\documentclass{aa}    
\usepackage{txfonts}
\usepackage{longtable}  
\usepackage{natbib}
\usepackage{graphicx}

\bibpunct[, ]{(}{)}{,}{a}{}{,}

\usepackage{color}

\newcommand{\Teff}{T_{\rm eff}}
\newcommand{\eps}[1]{\log\varepsilon_{\rm #1}}
\newcommand{\kms}{km\,s$^{-1}$}
\newcommand{\iso}[2]{\mbox{$^{#1}{\rm #2}$}}
\newcommand{\kH}{$S_{\!\!\rm H}$}    
\newcommand{\Eexc}{$E_{\rm exc}$}
\newcommand{\paperI}{\citet[][]{2010A&A...524A..58T}}
\newcommand{\paperII}{\citet[][]{2015A&A...583A..67J}}

\begin{document}

\title{The formation of the Milky Way halo and its dwarf satellites; a
  NLTE-1D abundance analysis. II. Early chemical enrichment 
    }

\author{
  L. Mashonkina\inst{1,2} \and
  P. Jablonka\inst{3,4} \and
  T. Sitnova\inst{2} \and
  Yu. Pakhomov\inst{2} \and
  P. North\inst{3}
}
 
\offprints{L. Mashonkina; \email{lima@inasan.ru}}
\institute{
     Universit\"ats-Sternwarte M\"unchen, Scheinerstr. 1, D-81679 M\"unchen, 
     Germany \\ \email{lyuda@usm.lmu.de}
\and Institute of Astronomy, Russian Academy of Sciences, RU-119017 Moscow, 
     Russia \\ \email{lima@inasan.ru}
\and Laboratoire d' Astrophysique, Ecole Polytechnique F\'ed\'erale de Lausanne (EPFL), Observatoire de Sauverny, CH-1290 Versoix, Switzerland 
\and GEPI, Observatoire de Paris, CNRS, Universit\'e Paris Diderot, F-92125 Meudon Cedex, France
}

\date{Received  / Accepted }

\abstract{
We present the non-local thermodynamic equilibrium (NLTE) abundances of up to 10 chemical species in a sample of 59 very metal-poor (VMP, $-4 \le$ [Fe/H] $\precsim -2$) stars in seven dwarf spheroidal galaxies (dSphs) and in the Milky Way (MW) halo.
Our results are based on high-resolution spectroscopic datasets and homogeneous and accurate atmospheric parameters determined in Paper~I.

We show that once the NLTE effects are properly taken into account, all massive
  galaxies in our sample, that is, the MW halo and the classical dSphs Sculptor, Ursa Minor, Sextans, and Fornax, reveal a similar plateau at [$\alpha$/Fe] $\simeq$ 0.3 for each of the $\alpha$-process elements: Mg, Ca, and Ti. 
   We put on a firm ground the evidence for a decline in $\alpha$/Fe with increasing metallicity in the Bo\"otes~I ultra-faint dwarf galaxy (UFD), that is most probably due to the ejecta of type Ia supernovae.
For Na/Fe, Na/Mg, and Al/Mg, the MW halo and all dSphs reveal indistinguishable trends with metallicity, suggesting that the processes of Na and Al synthesis are identical in all systems, independent of their mass.
The dichotomy in the [Sr/Ba] versus [Ba/H] diagram is observed in the classical dSphs, similarly to  the MW halo, calling for two different nucleosynthesis channels for Sr.
We show that Sr in the massive galaxies is well correlated with Mg suggesting a strong link to massive stars and that  
its production is essentially independent of Ba, for most of the [Ba/H] range.
Our three UFDs, that is Bo\"otes~I, UMa~II, and Leo~IV, are 
depleted in Sr and Ba relative to Fe and Mg, with very similar ratios of [Sr/Mg] $\simeq -1.3$ and [Ba/Mg] $\simeq -1$ on the entire range of their Mg abundances.
The subsolar Sr/Ba ratios of Bo\"otes~I and UMa~II indicate a common r-process origin of their  neutron-capture elements. Sculptor remains the classical dSph, in which the evidence for inhomogeneous mixing in the early evolution stage, at [Fe/H] $< -2$, is the strongest.  
}


\titlerunning{NLTE abundances of very metal-poor stars in the dwarf spheroidal galaxies}
\authorrunning{Mashonkina et al.}

\maketitle

%
\section{Introduction}\label{Sect:intro}

We aim at understanding the physical conditions at the onset of star
formation in galaxies. The stellar abundance trends and dispersions of
the most metal-poor stars reveal the nature of the first generations of stars
(e.g., mass, numbers, spatial distribution), and the level of
homogeneity of the primitive interstellar medium (e.g., size/mass of
star forming regions, nature and energetics of the explosion of
supernovae). The proximity of the Local Group dwarf spheroidal
galaxies (dSphs) allows the derivation of chemical abundances with
comparable quality as in the Milky Way (MW). The comparison of these
galaxies with very different masses, star formation histories, and
level of chemical enrichment can bring crucial information on the
universality of the physical processes at play. This type of
comparative analyses is at the frontier fields between stellar
and galaxy evolution, as for example, it provides a very detailed
testbed for the $\Lambda$CDM paradigm, improves our understanding of
fundamental parameters, such as the sampling of the initial mass
function, but also provides most wanted constraints on the
nucleosynthetic origin of still puzzling categories of chemical species,
such as the neutron-capture elements or the odd-nuclear charge (odd-$Z$) elements
sodium and aluminum.

As a consequence of their distance, most of the stars accessible in
dSphs at high spectral resolution are giants. Low total gas
pressure and low electron number density lead to departures from local
thermodynamical equilibrium (LTE) in their atmospheres, while the
non-local thermodynamical equilibrium (NLTE) line formation impacts
each chemical species differently, in magnitude and sign, depending on
the stellar atmosphere parameters and element abundances \citep[for review,
  see][]{Asplund2005ARAA,Mashonkina_review2013}. Therefore, ignoring
the NLTE effects for stellar samples covering broad metallicity ranges
can lead to a distorted picture of the galactic abundance trends.

This paper is the second part of a project aiming at providing a homogeneous set
of NLTE elemental abundances for the very and extremely metal-poor (VMP: [Fe/H]\footnote{In
  the classical notation, where [X/H] = $\log(N_{\rm X}/N_{\rm
    H})_{star} - \log(N_{\rm X}/N_{\rm H})_{Sun}$.} $\le -2$, EMP: [Fe/H] $\le -3$) stars both in the Local Group dwarf spheroidal galaxies and
in the halo of the Milky Way. Accurate atmospheric
parameters for our stellar sample were derived in our previous study \citep[][hereafter, Paper~I]{dsph_parameters}.

Our work has been preceded by a number of important efforts to evaluate the
NLTE effects on determination of 
chemical abundances for late-type stars.


For the chemical species of our interest in the
metallicity range similar to that of our study, from extremely to very
metal-poor stars, the NLTE abundance corrections  were calculated for \ion{Na}{i}
\citep{1998A&A...338..637B,2000ARep...44..790M,2007A&A...464.1081A}, \ion{Mg}{i}
\citep{Thevenin2000,2000ARep...44..530S,2000A&A...362.1077Z,2010A&A...509A..88A},
\ion{Al}{i} \citep{1997A&A...325.1088B,2008A&A...481..481A}, \ion{Si}{i}
\citep{2009A&A...503..533S}, \ion{Ca}{i}
\citep{Thevenin2000,mash_ca,2012A&A...541A.143S}, \ion{Ti}{i}-\ion{Ti}{ii}
\citep{2011MNRAS.413.2184B,2016AstL...42..734S}, \ion{Sr}{ii}
\citep{Mashonkina2001sr,2011A&A...530A.105A,2012A&A...546A..90B}, and
\ion{Ba}{ii} \citep{Mashonkina1999,2009A&A...494.1083A}.  Recently,
\citet{lick_paperII} derived the NLTE abundances of a wealth of
elements, from Li to Eu, for a sample of 51 Galactic dwarf
stars covering the $-2.6 \le$ [Fe/H] $\le 0.2$ range. They have the closest methodology
to the present study, and are complementary in the spectral type of stars as
well as the metallicity range.

Since in most cases, the departures from LTE for individual lines
depend strongly on the stellar atmosphere parameters and elemental
abundances, we do not use pre-existing NLTE abundance corrections from
the literature. Instead, the NLTE calculations are performed for each
star and chemical species. Our approach stands out among the previous
publications by three specific features at least: i) the [X/Fe] abundance
ratios are calculated by correcting the iron abundances from
NLTE effects as well, unlike most of our predecessors; ii) the NLTE
abundances are calculated in idem homogeneous way for all
elements; iii) we go beyond the Milky Way stellar population and
address the chemical trends of the Local Group dwarf spheroidal
galaxies.

In the following, Sect.~\ref{Sect:basics} briefly describes our
stellar sample, the observational material, and the adopted method to
derive the atmospheric parameters. Details of the NLTE calculations
and the abundance determinations, including atomic models, line list,
and codes, are provided in
Sect.~\ref{Sect:Method}. Section~\ref{Sect:abundances} details the
impact of the NLTE treatment on each individual chemical species,
while Sect.~\ref{Sect:dsph_halo} presents the abundance trends within
each galaxy and draws comparisons between them to infer their chemical
evolution. It also provides new constrains on the nucleosynthetic
origin of the neutron-capture elements. Our results are summarized in
Sect.\,\ref{Sect:Conclusions}.

\section{Stellar sample, observational material, and atmospheric parameters}\label{Sect:basics}

Our stellar sample, the observational material, and the
determination of atmospheric parameters were described in Paper~I. We summarize below
the main features. 

{\it Sample.} We are working with two stellar samples of cool giants in
  the $-4 \lesssim$ [Fe/H] $\lesssim -2$ metallicity range. One is composed of
  23 stars in the Milky Way halo, the other encompasses 36 stars in a number of
  MW satellites, namely, the classical dSphs Sculptor (Scl), Ursa Minor (UMi),
  Fornax (Fnx), and Sextans (Sex) and the ultra-faint dwarfs (UFDs) Bo\"otes~I,
  UMa~II, and Leo~IV.

{\it Material.} Each star was observed at high spectral resolution, R =
  $\lambda/\Delta\lambda \ge$ 25\,000. We used the stellar spectra from  public
  and private archives as well as published equivalent widths.
  
{\it Stellar atmosphere parameters.} We used a combination of  photometric and
  spectroscopic methods  to derive a homogeneous set of stellar
  atmosphere parameters: effective temperature $\Teff$, surface gravity log~$g$,
  iron abundance [Fe/H], and microturbulence velocity $\xi_t$. Our spectroscopic
  analyses take advantage of NLTE line formation.

  For both stellar samples we rely on photometric effective temperatures. The
  surface gravities of the dSph stars were calculated by applying the standard
  relation between log~$g$, $\Teff$, the absolute bolometric magnitude, and the
  stellar mass, that was assumed to be 0.8\,M$_\odot$. The \ion{Fe}{i}- and
  \ion{Fe}{ii}-based NLTE abundances of the dSph stars are fully consistent in the [Fe/H]
  $> -3.7$ regime. We therefore applied the \ion{Fe}{i}/\ion{Fe}{ii} ionisation
  equilibrium method to determine the surface gravities of the MW giants, for
  which distances are mostly unknown or inaccurate. The abundance difference between
  \ion{Fe}{i} and \ion{Fe}{ii} does not exceed 0.12~dex and 0.06~dex
  for the dSph and MW stars, respectively.

  We noted that the NLTE treatment fails to achieve consistent abundances from \ion{Fe}{i} and
  \ion{Fe}{ii} in the most metal-poor ([Fe/H] $< -3.7$) stars in the dSphs. In consequence, we did
  not enforce the \ion{Fe}{i}/\ion{Fe}{ii} ionisation equilibrium for the most
  metal-poor star in the MW comparison sample, HE1357-0123 ([Fe/H] $\simeq
  -3.9$), but used an upward revised, by 0.2~dex, log~$g$ derived by
  \citet{Cohen2013} from the isochrone method.

  The problem of the \ion{Fe}{i}/\ion{Fe}{ii} ionisation equilibrium of our most
  metal-poor stars most probably relates to the uncertainty in the photometric
  calibration of $\Teff$ at these extremely low metallicities. Hence, our final
  [Fe/H] values, for the entire sample, are based on lines of \ion{Fe}{ii} only.
 
  Our sample stars and their atmospheric parameters are listed in Table\,\ref{Tab:parameters}. 

\begin{table} 
 \caption{\label{Tab:parameters} Atmospheric parameters of the investigated sample.} 
 \centering
 \begin{tabular}{lclcc}
\hline\hline \noalign{\smallskip}
ID & $\Teff$[K] &  log~$g$ &  [Fe/H] &  $\xi_t$[\kms]  \\
\noalign{\smallskip} \hline \noalign{\smallskip}
Scl ET0381     & 4570 & 1.17 &  $-2.19$ & 1.7  \\
Scl002\_06     & 4390 & 0.68 &  $-3.11$ & 2.3  \\
Scl03\_059     & 4530 & 1.08 &  $-2.88$ & 1.9  \\
Scl031\_11     & 4670 & 1.13 &  $-3.69$ & 2.0  \\
Scl074\_02     & 4680 & 1.23 &  $-3.06$ & 2.0  \\
Scl07-49       & 4630 & 1.28 &  $-2.99$ & 2.0  \\
Scl07-50       & 4800 & 1.56 &  $-4.00$ & 2.2  \\
Scl11\_1\_4296 & 4810 & 1.76 &  $-3.70$ & 1.9  \\
Scl6\_6\_402   & 4890 & 1.78 &  $-3.66$ & 1.8  \\
Scl S1020549   & 4650 & 1.35 &  $-3.67$ & 2.0  \\
Scl1019417     & 4280 & 0.50 &  $-2.48$ & 2.0  \\		 
Fnx05-42       & 4350 & 0.70 &  $-3.37$ & 2.3  \\
Sex11-04       & 4380 & 0.57 &  $-2.60$ & 2.2 \\
Sex24-72       & 4400 & 0.76 &  $-2.84$ & 2.2 \\
UMi396     & 4320 & 0.70 & $-2.26$ & 2.5 \\	   
UMi446     & 4600 & 1.37 & $-2.52$ & 2.5 \\
UMi718     & 4630 & 1.13 & $-2.00$ & 2.0 \\
UMi COS233 & 4370 & 0.77 & $-2.23$ & 2.0 \\
UMi JI19   & 4530 & 1.00 & $-3.02$ & 2.0 \\ 
UMi20103   & 4780 & 1.55 & $-3.09$ & 2.0 \\
UMi28104   & 4275 & 0.65 & $-2.12$ & 2.0 \\
UMi33533   & 4430 & 0.75 & $-3.14$ & 2.0 \\
UMi36886   & 4400 & 0.82 & $-2.56$ & 2.0 \\
UMi41065   & 4350 & 0.63 & $-2.48$ & 2.0 \\
Boo-033    & 4730 & 1.4  & $-2.26$ & 2.3  \\
Boo-041    & 4750 & 1.6  & $-1.54$ & 2.0  \\
Boo-094    & 4570 & 1.01 & $-2.69$ & 2.2  \\
Boo-117    & 4700 & 1.4  & $-2.09$ & 2.3  \\
Boo-127    & 4670 & 1.4  & $-1.93$ & 2.3  \\
Boo-130    & 4730 & 1.4  & $-2.20$ & 2.3  \\
Boo-980    & 4760 & 1.8  & $-2.94$ & 1.8  \\ 
Boo-1137   & 4700 & 1.39 & $-3.76$ & 1.9  \\
UMa~II-S1  & 4850 & 2.05 & $-2.96$ & 1.8  \\
UMa~II-S2  & 4780 & 1.83 & $-2.94$ & 2.0  \\
UMa~II-S3  & 4560 & 1.34 & $-2.26$ & 1.8  \\
Leo~IV-S1  & 4530 & 1.09 & $-2.58$ & 2.2  \\
HD~2796     & 4880 & 1.55 & $-2.32$ & 1.8 \\
HD~4306     & 4960 & 2.18 & $-2.74$ & 1.3 \\
HD~8724     & 4560 & 1.29 & $-1.76$ & 1.5 \\
HD~108317   & 5270 & 2.96 & $-2.18$ & 1.2 \\
HD~122563   & 4600 & 1.32 & $-2.63$ & 1.7  \\
HD~128279   & 5200 & 3.00 & $-2.19$ & 1.1  \\
HD~218857   & 5060 & 2.53 & $-1.92$ & 1.4  \\
HE0011-0035 & 4950 & 2.0  & $-3.04$ & 2.0 \\
HE0039-4154 & 4780 & 1.6  & $-3.26$ & 2.0 \\ 
HE0048-0611 & 5180 & 2.7  & $-2.69$ & 1.7  \\
HE0122-1616 & 5200 & 2.65 & $-2.85$ & 1.8  \\
HE0332-1007 & 4750 & 1.5  & $-2.89$ & 2.0  \\
HE0445-2339 & 5165 & 2.2  & $-2.76$ & 1.9  \\
HE1356-0622 & 4945 & 2.0  & $-3.45$ & 2.0  \\
HE1357-0123 & 4600 & 1.2  & $-3.92$ & 2.1  \\
HE1416-1032 & 5000 & 2.0  & $-3.23$ & 2.1  \\
HE2244-2116 & 5230 & 2.8  & $-2.40$ & 1.7  \\
HE2249-1704 & 4590 & 1.2  & $-2.94$ & 2.0  \\
HE2252-4225 & 4750 & 1.55 & $-2.76$ & 1.9  \\
HE2327-5642 & 5050 & 2.20 & $-2.92$ & 1.7  \\
BD $-11^\circ$ 0145 & 4900 & 1.73 & $-2.18$ & 1.8 \\
CD $-24^\circ$ 1782 & 5140 & 2.62 & $-2.72$ & 1.2 \\
BS16550-087         & 4750 & 1.5  & $-3.33$ & 2.0  \\
\noalign{\smallskip}\hline \noalign{\smallskip}
\end{tabular}
\end{table}

\section{Method}\label{Sect:Method}

\subsection{NLTE calculations}

The  present investigation is based on the NLTE methods developed in our
earlier studies and documented in a number of papers (see Table~\ref{Tab:nlte} for the refereces), in which the atomic data and the
questions of line formation have been considered in detail.
We have updated collisional data for a number of chemical species. For \ion{Sr}{ii} we apply here the electron-impact excitation rate
coefficients from {\it ab initio} calculations of \citet{2002MNRAS.331..875B}.  For
\ion{Al}{i}, \ion{Si}{i}, and \ion{Ca}{i}, the inelastic collisions with neutral
hydrogen particles are treated using the accurate rate coefficients from
quantum-mechanical calculations of \citet{Belyaev2013_Al,Belyaev2014_Si}, and
\citet{Belyaev2016_Ca}, respectively. For \ion{Ti}{i-ii}, \ion{Fe}{i-ii},
\ion{Sr}{ii}, and \ion{Ba}{ii} we rely on the \citet{Drawin1968} approximation, as
implemented by \citet{Steenbock1984}, and apply a scaling factor \kH\ to the
Drawinian rates. The magnitude of \kH\ indicated in Table~\ref{Tab:nlte} was
estimated empirically for every chemical species in our earlier papers.

\begin{table} 
 \caption{\label{Tab:nlte} Atomic models used in this study.}
 \centering
 \begin{tabular}{lll}\hline\hline \noalign{\smallskip}
 Species & Reference & \ion{H}{i} collisions  \\
\noalign{\smallskip} \hline \noalign{\smallskip}
\ion{Na}{i}    &   \citet{alexeeva_na} &  BBD10 \\
\ion{Mg}{i}    &   \citet{mash_mg13}   &  BBS12 \\
\ion{Al}{i}    &   \citet{Baumueller_al1} & B13  \\
\ion{Si}{i}    &   \citet{Shi_si_sun}  &  BYB14 \\
\ion{Ca}{i-ii} &   \citet{mash_ca}     &  BYG16 \\
\ion{Ti}{i-ii} &   \citet{sitnova_ti}    &  \kH\,= 1 \\
\ion{Fe}{i-ii} &   \citet{mash_fe}       &   \kH\,= 0.5 \\
\ion{Sr}{ii}   &   \citet{1997ARep...41..530B} &\kH\,= 0.01 \\
\ion{Ba}{ii}   &   \citet{Mashonkina1999}& \kH\,= 0.01 \\
\ion{Eu}{ii}   &   \citet{mash_eu}       & \kH\,= 0.1 \\
\noalign{\smallskip}\hline \noalign{\smallskip}
\multicolumn{3}{l}{{\bf Notes.} Collisions with \ion{H}{i} are treated following to } \\ 
\multicolumn{3}{l}{BBD10 = \citet{barklem2010_na},  } \\
\multicolumn{3}{l}{BBS12 = \citet{mg_hyd2012}, } \\
\multicolumn{3}{l}{B13 = \citet{Belyaev2013_Al}, BYB14 = \citet{Belyaev2014_Si},} \\
\multicolumn{3}{l}{BYG16 = \citet{Belyaev2016_Ca}, } \\
\multicolumn{3}{l}{\citet{Steenbock1984}, with a scaling factor of \kH.} \\
\end{tabular}
\end{table}

To solve the coupled radiative transfer and statistical equilibrium
(SE) equations, we employed a revised version of the {\sc DETAIL} code
\citep{detail} based on the accelerated lambda iteration method, as
described in \citet{rh91,rh92}. An update of the opacity package in {\sc DETAIL} was presented by
\citet{mash_fe}. 

\subsection{Line selection, atomic data, and notes on abundance determinations}\label{Sect:linelist}

The spectral lines used in the abundance analysis are listed in
  Table\,\ref{Tab:all_lines} together with their atomic
  parameters. The $gf$-values are taken from VALD3, with the exception of \ion{Fe}{ii}, for which we used $gf$-values from \citet{RU} that were corrected by $+0.11$~dex, following the
  recommendation of \citet{Grevesse1999}.
The van der Waals broadening was accounted for by applying accurate data based on
the perturbation theory \citep[see][and references therein]{2000A&AS..142..467B}, with
 the exception of \ion{Ca}{i}, for which we employed the van der Waals damping constants, $\Gamma_6$, based on
laboratory measurements of \citet{1981A&A...103..351S}, and a few selected lines of
\ion{Na}{i}, \ion{Al}{i}, \ion{Mg}{i}, and \ion{Ba}{ii}, for which $\Gamma_6$ was
estimated empirically from solar line-profile fitting by \citet[][\ion{Na}{i} and
  \ion{Al}{i}]{mg_c6}, \citet[][\ion{Mg}{i}]{mash_mg13}, and \citet[][\ion{Ba}{ii}]{Mashonkina2008}.

In VMP stars, aluminum can only be observed in the resonance doublet lines \ion{Al}{i} 3944 and 3961\,\AA\ located in the spectral range that is crowded even at so low metallicity and rather noisy, in particular, in case of the dSph stars. The \ion{Al}{i} 3944.006\,\AA\ line is heavily blended by the molecular CH 3943.85 and 3944.16\,\AA\ lines, and it was only used for an EMP star Scl031\_11, which has low carbon abundance \citep{2015A&A...583A..67J} and the best-quality observed spectrum. The second line, \ion{Al}{i} 3961\,\AA, lies in the wing of \ion{Ca}{ii} 3968\,\AA\ and must be synthesised with a pre-determined calcium abundance.

For the Si abundance determination, we prefer to employ \ion{Si}{i} 4102\,\AA, which lies in the far wing of H$_\delta$, but not affected by any other atomic or molecular lines. However, for 9 stars in the dSphs and 12 MW stars with the best-quality spectra we use also \ion{Si}{i} 3905.52\,\AA. This line is blended by the molecular CH 3905.68\,\AA\ line, and the Si abundance was derived either via spectral synthesis (six dSph stars) or by estimating 
a contribution of the CH line to the 3905\,\AA\ blend via spectral synthesis. The C/Fe abundance ratios were taken from \citet{2010ApJ...719..931C}, \citet{Norris2010}, \citet{2010A&A...524A..58T}, \citet{Cohen2013}, and \citet{2015A&A...583A..67J}. Our analysis of the CH 4310\,\AA\ band in the available spectra has confirmed the literature data on [C/Fe]. It is worth noting that the investigated dSph stars have low carbon abundance, with [C/Fe] $\le 0$. The exception is Boo-1137, with [C/Fe] = 0.25 \citep{Norris2010}. We evaluated the difference in the Si abundance derived without and with the CH line taken into account. For Boo-1137 it amounts to 0.32~dex. The stars of the MW comparison sample were preselected by requiring [C/Fe] $\le 0$. 

The best observed neutron-capture elements are strontium and
barium. Unfortunately, even their abundances are not always accessible in our MW
and dSph sample stars. This is partly because the observations did not cover the
strongest lines of these elements located in the blue spectral range and partly
because some spectra are too noisy in these regions. For example, \ion{Sr}{ii}
4077, 4215\,\AA\ are missing in the Bo\"otes~I spectra of \citet{Gilmore2013}
and so are the \ion{Sr}{ii} and \ion{Ba}{ii} resonance lines in Scl07-49,
HD~218857, and CD$-24^\circ$1782. For several stars, the resonance lines of
\ion{Sr}{ii} and/or \ion{Ba}{ii} could not be extracted from noise. This is the case of
five stars in the Sculptor dSph, that is, 11\_1\_4296 and S1020549 from
\citet{2015ApJ...802...93S} and 002\_06, 031\_11, and 074\_02 from
\citet{2015A&A...583A..67J}, and a MW halo star HE1357-0123 from
\citet{Cohen2013}. 
We caution that we could not use the \ion{Ba}{ii} equivalent widths of
\citet{2015MNRAS.449..761U} as a consequence of strong inconsistencies between
lines. For example, in UMi718, \citet{2015MNRAS.449..761U} measured $W_{obs}$ =
63.5\,m\AA\ for \ion{Ba}{ii} 5853\,\AA\ (\Eexc\ = 0.6~eV, log$gf$ = $-1$), while
$W_{obs}$ = 28.8\,m\AA\ for \ion{Ba}{ii} 6496\,\AA\ (\Eexc\ = 0.6~eV, log$gf$ =
$-0.38$). For \ion{Ba}{ii} 4554\,\AA\ (\Eexc\ = 0, log$gf$ = 0.17) in UMi446, they
give $W_{obs}$ = 72.1\,m\AA, but larger value of $W_{obs}$ = 144.4\,m\AA\ for
\ion{Ba}{ii} 6496\,\AA.

For \ion{Sr}{ii}, we attempted to use both resonance lines when they are available. The \ion{Sr}{ii} 4215.539\,\AA\ line is notably
  blended by \ion{Fe}{i} 4215.426\,\AA\ in the [Fe/H] $> -3$ stars and therefore was ignored in the
  abundance analysis, if \ion{Sr}{ii} 4077\,\AA\ was available. The case of 
  Boo-127 ([Fe/H] = $-1.93$) makes an exception to this rule. As in \citet{2016ApJ...826..110F}, 
 the synthetic spectrum analysis results in a substantially lower abundance from \ion{Sr}{ii} 4077\,\AA, by 0.5~dex, compared with that from \ion{Sr}{ii} 4215\,\AA. The Sr abundance of Boo-127 was derived from \ion{Sr}{ii} 4215\,\AA.

The \ion{Ba}{ii} 4934\,\AA\ and 6141\,\AA\ lines are blended with some
\ion{Fe}{i} lines \citep[see details in][]{2015A&A...583A..67J}. Whenever the spectrum was
available, the Ba abundance was determined from spectral synthesis. Otherwise, the
contribution of the \ion{Fe}{i} lines to the blends at 4934\,\AA\ and
6141\,\AA\ was estimated by computing their synthetic spectra. When the blending
between the Fe and Ba lines was substantial, then \ion{Ba}{ii} 6141\,\AA\ was not used.

The lines of \ion{Sr}{ii}, \ion{Ba}{ii}, and \ion{Eu}{ii} are composed of multiple components because each of these chemical elements is represented by several isotopes. For \ion{Sr}{ii} 4077 and 4215\,\AA, their isotopic splitting (IS) and hyper-fine splitting (HFS) structure is taken into account using the atomic data from \citet{1983HyInt..15..177B} and the solar system Sr isotope abundance ratios from \citet{Lodders2009}. It is worth noting that the difference between the solar system and the r-process Sr isotope mixture \citep{Arlandini1999} produces negligible effect on the derived stellar Sr abundances.

In contrast, the Ba abundance derived from the 4554\,\AA\ and 4934\,\AA\ resonance lines depends on the isotope mixture
adopted in the calculations because of substantial separations of 57 and 78~m\AA, respectively, between the HFS components \citep{Brix1952,1980ZPhyA.295..311S,1982PhRvA..25.1476B,1983ZPhyA.311...41B}. Since the odd-$A$ isotopes \iso{135}{Ba} and
\iso{137}{Ba} have very similar HFS, the abundance is essentially dependent on the
total fractional abundance of these odd isotopes, $f_{\rm odd}$. The larger $f_{\rm odd}$, the broader the resonance line is, and the larger energy it absorbs. In the Solar system matter,
$f_{\rm odd}$ = 0.18 \citep{Lodders2009}, however, larger value of $f_{\rm odd}$ =
0.438 \citep{Kratz2007} to 0.72 \citep{McWilliam1998} is predicted for pure
r-process production of barium.
We inspected the influence of the $f_{\rm odd}$ variation on the Ba abundance
determinations. 
In the four MW stars, with $W_{obs}$(4554\,\AA) $\le$ 33\,m\AA, the abundance derived from the \ion{Ba}{ii} resonance line is, in fact, insensitive to the adopted $f_{\rm odd}$ value. In Boo-1137, with $W_{obs}$(4934\,\AA) = 40.5\,m\AA, and Leo~IV-S1, with $W_{obs}$(4554\,\AA) = 62\,m\AA, the abundance
difference between using $f_{\rm odd}$ = 0.18 and 0.46 amounts to $\Delta\eps{}$ = 0.03~dex and 0.05~dex, respectively.  
A notable shift of $\Delta\eps{}$ = 0.18~dex was only found for HE0011-0035, with
$W_{obs}$(4554\,\AA) = 104\,m\AA.

Fortunately, the \ion{Ba}{ii}
5853, 6141, and 6497\,\AA\ subordinate lines are almost free of HFS effects. 
Therefore, in case the
subordinate lines are available, they were used to derive the Ba abundance, at the
exception of Boo-1137, for which \ion{Ba}{ii} 6141 and 6497\,\AA\ are
rather weak. No subordinate line could be measured for Leo~IV-S1, HE0011-0035,
HE0122-1616, HE1356-0622, HE1416-1032, HE2249-1704, and BS16550-087, and the Ba abundance was determined from the resonance lines, adopting $f_{\rm odd}$ = 0.46, as predicted by \citet{Travaglio1999} for the r-process Ba isotope mixture.

We could determine the Eu abundances for six dSph stars and 12 MW halo
 stars. For eight of them, we work on the observed spectra and
 perform spectral synthesis of the \ion{Eu}{ii} lines. For the other stars, we use
 the equivalent widths published by \citet[][Scl~1019417]{2012AJ....144..168K},
 \citet[][UMi~28104, 33533, 36886, 41065, and JI19]{2010ApJ...719..931C}, and
 \citet[][four MW stars]{Cohen2013}. For all, we properly account for the
 HFS and IS structure thanks to the data of \citet{Lawler_Eu}.

\subsection{Codes and model ingredients}\label{Sect:codes}

In this study, the synthetic spectrum method is used to derive the element abundances from \ion{Al}{i} 3944, 3961\,\AA, \ion{Si}{i} 3905, 4102\,\AA, \ion{Sr}{ii} 4215\,\AA, and \ion{Ba}{ii} 6141\,\AA\ to take into account the blending lines and from \ion{Ba}{ii} 4934, 4554\,\AA\ to take into account HFS structure of the lines. 
For this we used the codes {\sc SIU}
\citep{Reetz} and {\sc synthV\_NLTE} \citep{Ryabchikova2015} that implement
 the pre-computed departure coefficients, $b_i = n_i^{\rm NLTE}/n_i^{\rm
  LTE}$, to calculate the NLTE line profiles for the NLTE species. Here, $n_i^{\rm
  NLTE}$ and $n_i^{\rm LTE}$ are the statistical equilibrium and thermal
(Saha-Boltzmann) number densities, respectively, from {\sc DETAIL}. The code {\sc
  synthV\_NLTE} is integrated within the IDL {\sc binmag3}\footnote{http://www.astro.uu.se/$\sim$oleg/download.html} code, written by
O. Kochukhov, finally allowing the user to determine the best fit to the observed
line profile. Line list for spectral synthesis has been extracted from the Vienna
Atomic Line Database\footnote{http://vald.astro.univie.ac.at/~vald3/php/vald.php}
\citep[VALD3,][]{2015PhyS...90e4005R}. Our test calculations with the solar model
atmosphere in a broad wavelength range (4209~\AA\ to 9111~\AA) have proved that
using two different codes, {\sc SIU} and {\sc synthV\_NLTE} + {\sc binmag3}, does
not produce systematic shifts in derived abundances, namely, the abundance
difference nowhere exceeds 0.03~dex.

 For all other lines we use their equivalent widths. For each line, we first
calculate the LTE abundance with the code {\sc WIDTH9}\footnote{\tt
  http://kurucz.harvard.edu/programs/WIDTH/} \citep[][modified by Vadim Tsymbal,
  private communication]{2005MSAIS...8...14K}. The NLTE abundance is then obtained by
applying the NLTE abundance correction, $\Delta_{\rm NLTE} =
\eps{NLTE}-\eps{LTE}$, computed with the code {\sc LINEC} \citep{Sakhibullin1983}
that uses the LTE and NLTE level populations from {\sc DETAIL}. 
Our test calculations with
{\sc LINEC} and {\sc WIDTH9} have proved that, for any given line, both codes lead to consistent LTE abundances within 0.01-0.02~dex.

In a similar homogeneous way, all the codes we used do treat continuum
scattering correctly; i.e., scattering is taken into account not only in the
absorption coefficient, but also in the source function.

As in Paper~I, we use the MARCS model structures \citep{Gustafssonetal:2008}.

\section{Stellar element abundances}\label{Sect:abundances}

The LTE and NLTE abundances of Na, Mg, Al, Si, Ca, Ti, Ni (only LTE), Sr, and Ba were
determined for each star, provided the corresponding lines could be measured.
The obtained results are available as online material (Table\,\ref{Tab:all_lines}).
Other chemical species measured in \paperII, namely, \ion{Sc}{ii},
\ion{Cr}{i}, \ion{Mn}{i}, \ion{Co}{i}, and \ion{Y}{ii}, are absent in our list,
because no NLTE calculations were performed in the literature for the atmospheric
parameters of our interest.

\begin{table*} 
 \caption{\label{Tab:all_lines} LTE and NLTE abundances from individual lines in the sample stars. This table is available in its entirety in a machine-readable
form in the online version. A portion is shown here for guidance
regarding its form and content. }
 \centering
 \begin{tabular}{lccrcrrr}\hline\hline \noalign{\smallskip}
 Atom & $\lambda$ & $E_{\rm exc}$ & $\log gf$ & $\log \Gamma_6/N_{\rm H}$ & EW & \multicolumn{2}{c}{$\eps{}$} \\
\cline{7-8}  \noalign{\smallskip}
      & (\AA)     & (eV)          &           & (rad/s$\cdot$cm$^3$) & (m\AA ) & LTE & NLTE \\
\noalign{\smallskip} \hline \noalign{\smallskip}
\multicolumn{8}{c}{ET0381} \\
Na I  & 5889.95 & 0.00 &  0.12   & $-$7.670 & 173.0 & 3.51 & 3.19 \\
Na I  & 5895.92 & 0.00 & $-$0.19 & $-$7.670 & 143.0 & 3.35 & 3.07 \\
Mg I  & 5172.68 & 2.71 & $-$0.45 & $-$7.267 & 197.2 & 5.04 & 5.06 \\
Mg I  & 5183.60 & 2.72 & $-$0.24 & $-$7.267 & 204.8 & 4.90 & 4.93 \\
Mg I  & 5528.41 & 4.35 & $-$0.50 & $-$7.180 &  59.5 & 4.99 & 5.00 \\
Al I  & 3961.52 & 0.01 & $-$0.34 & $-$7.315 & $-1$  & 2.72 & 2.82 \\
\noalign{\smallskip}\hline \noalign{\smallskip}
\multicolumn{8}{l}{{\bf Notes.} $\Gamma_6$ corresponds to 10\,000~K. The sources of the observed} \\ 
\multicolumn{8}{l}{equivalent widths, EWs, are indicated in Table\,\ref{Tab:AbundanceSummary}.} \\
\multicolumn{8}{l}{EW = $-1$ means using spectral synthesis.}
\end{tabular}
\end{table*}

\subsection{Impact of NLTE on derived chemical abundances}\label{Sect:nlte}

\begin{figure*} 
  \resizebox{90mm}{!}{\includegraphics{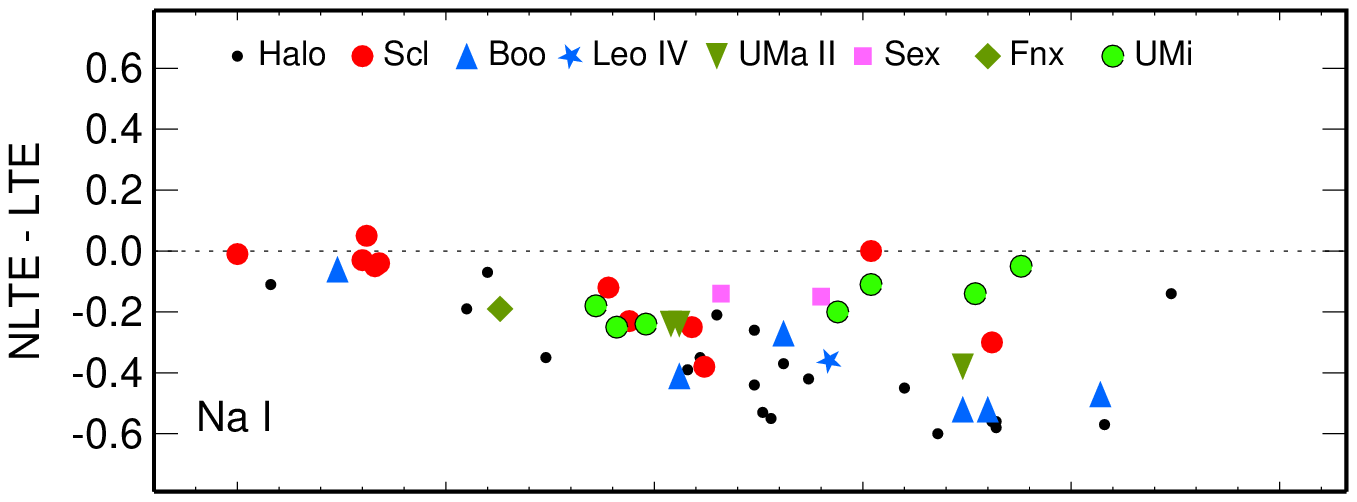}}
  \resizebox{90mm}{!}{\includegraphics{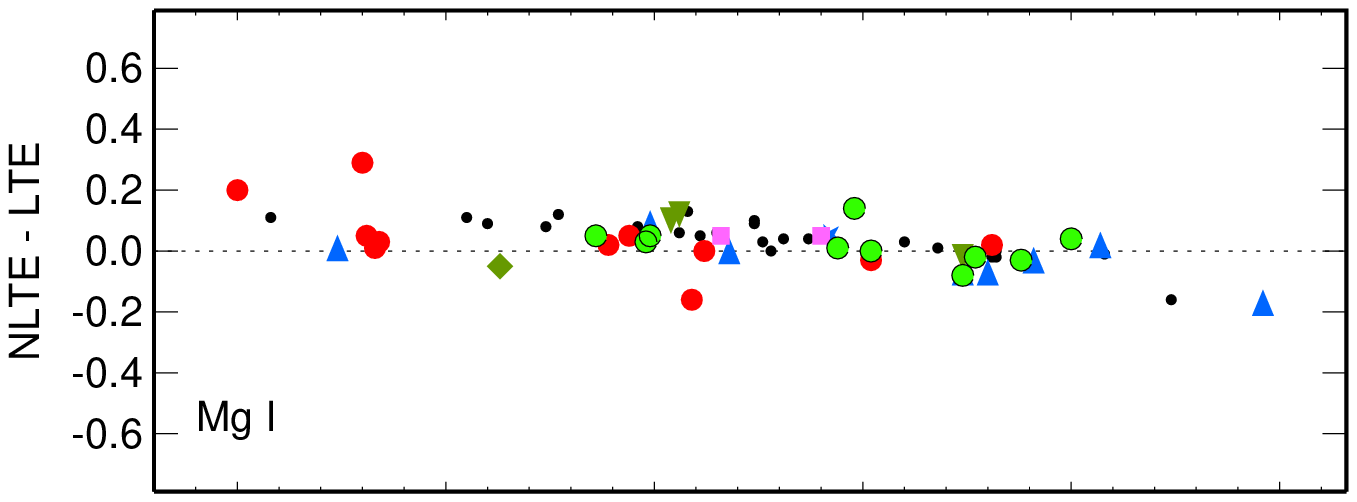}}

  \vspace{-13mm}  
  \resizebox{90mm}{!}{\includegraphics{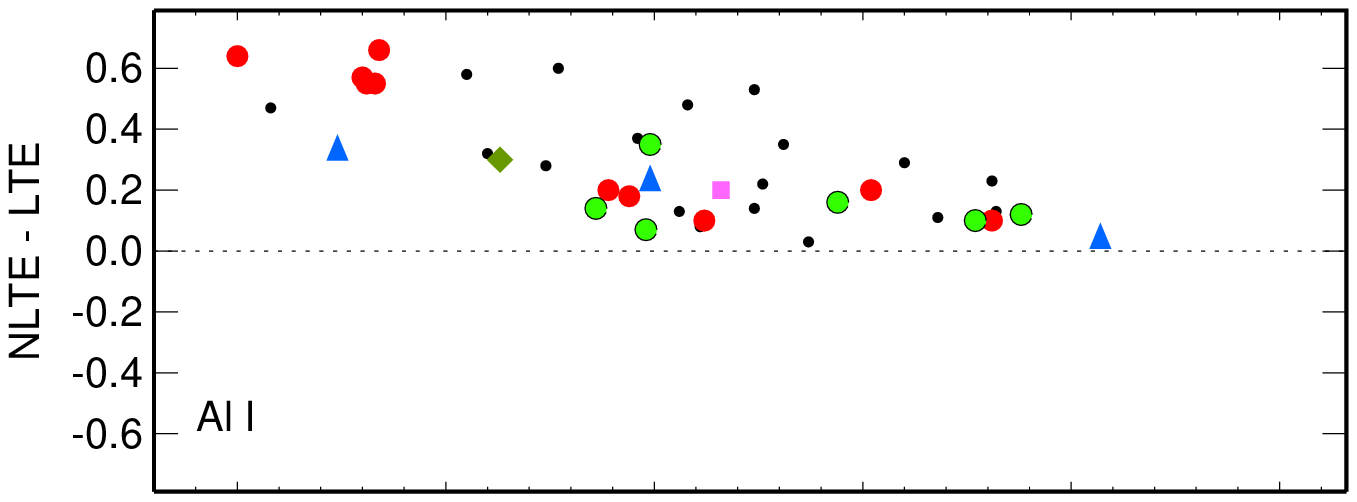}}
  \resizebox{90mm}{!}{\includegraphics{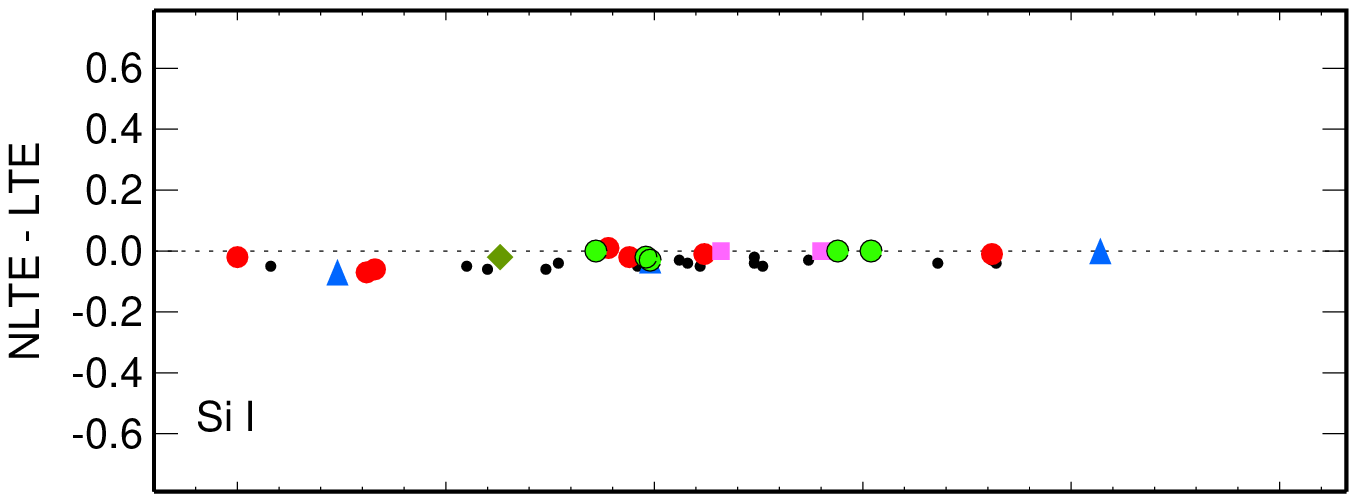}}
  
  \vspace{-13mm}  
  \resizebox{90mm}{!}{\includegraphics{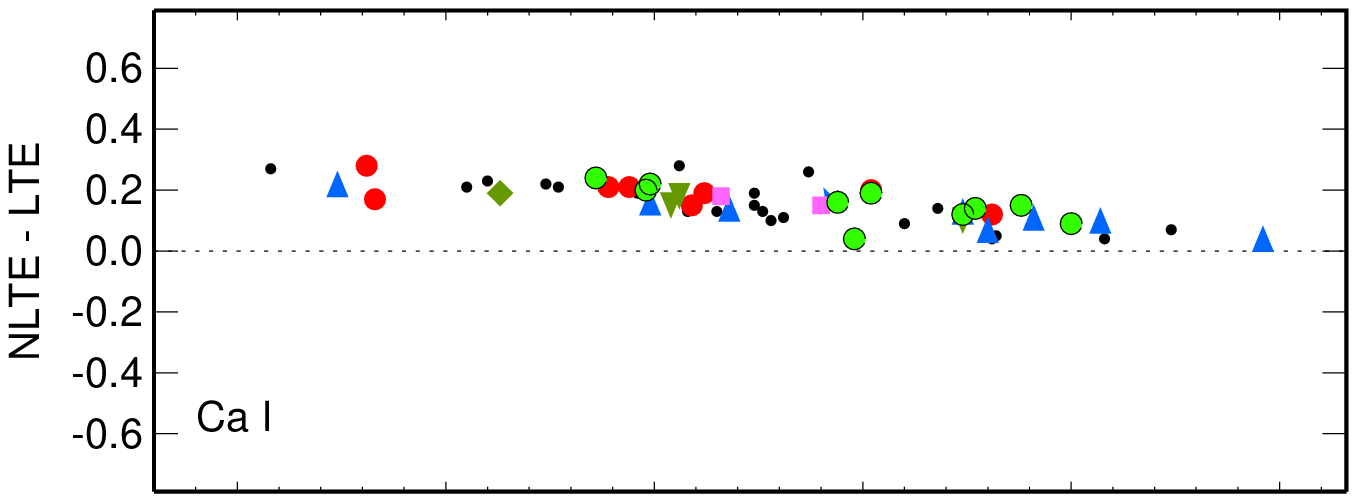}}
  \resizebox{90mm}{!}{\includegraphics{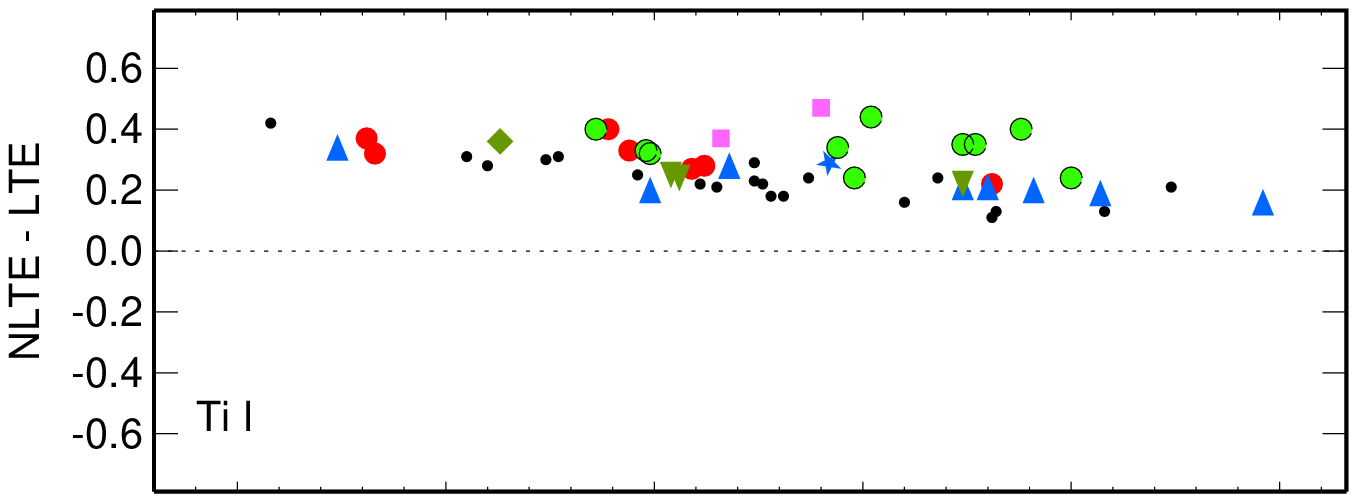}}
  
  \vspace{-13mm}  
  \resizebox{90mm}{!}{\includegraphics{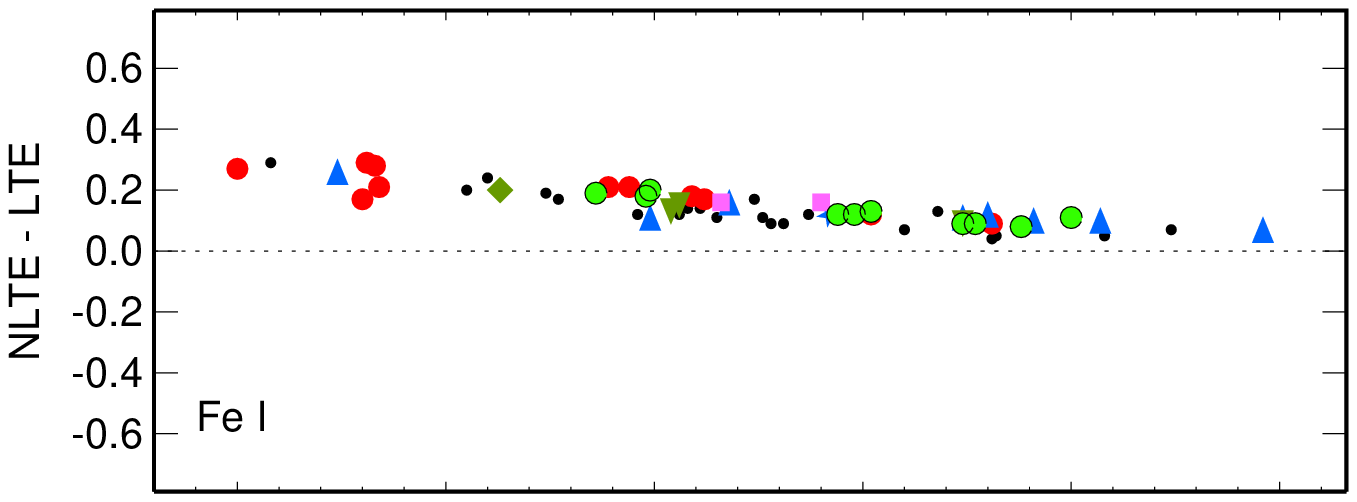}}
  \resizebox{90mm}{!}{\includegraphics{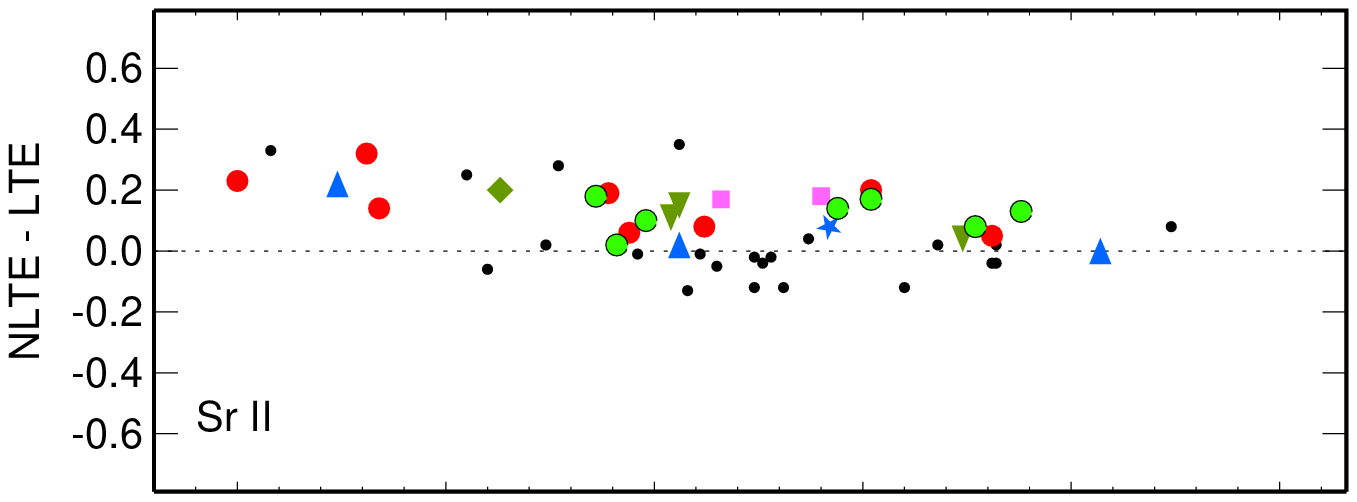}}

  \vspace{-13mm}  
  \resizebox{90mm}{!}{\includegraphics{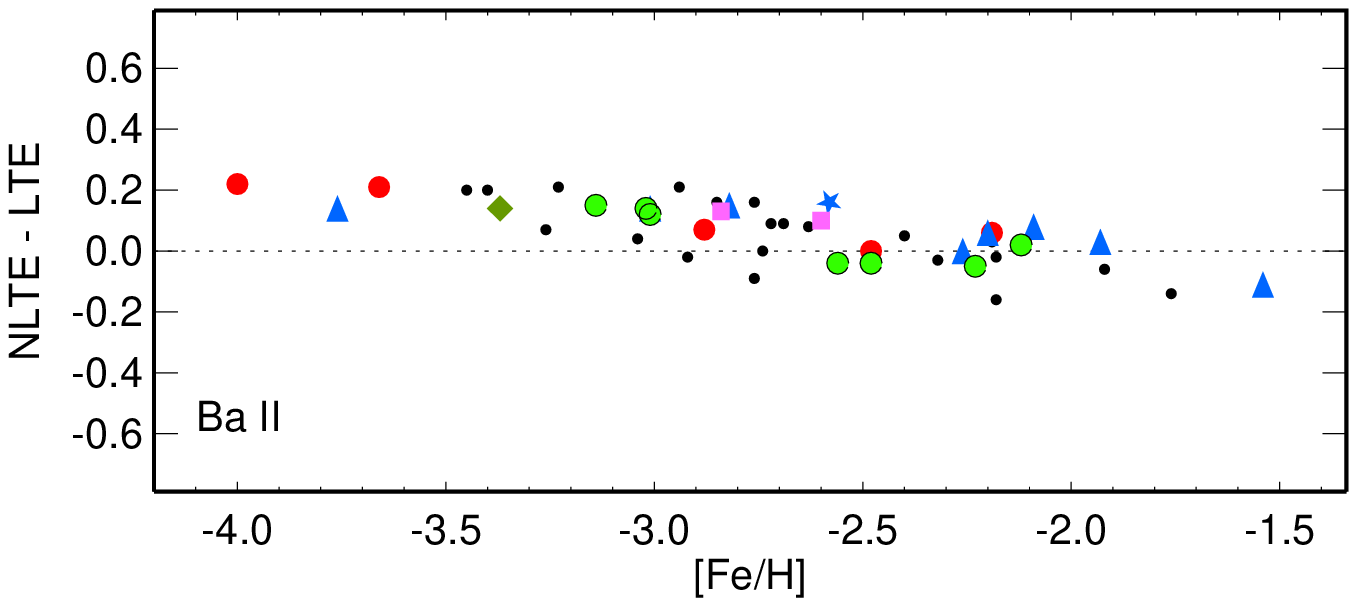}}
  \resizebox{90mm}{!}{\includegraphics{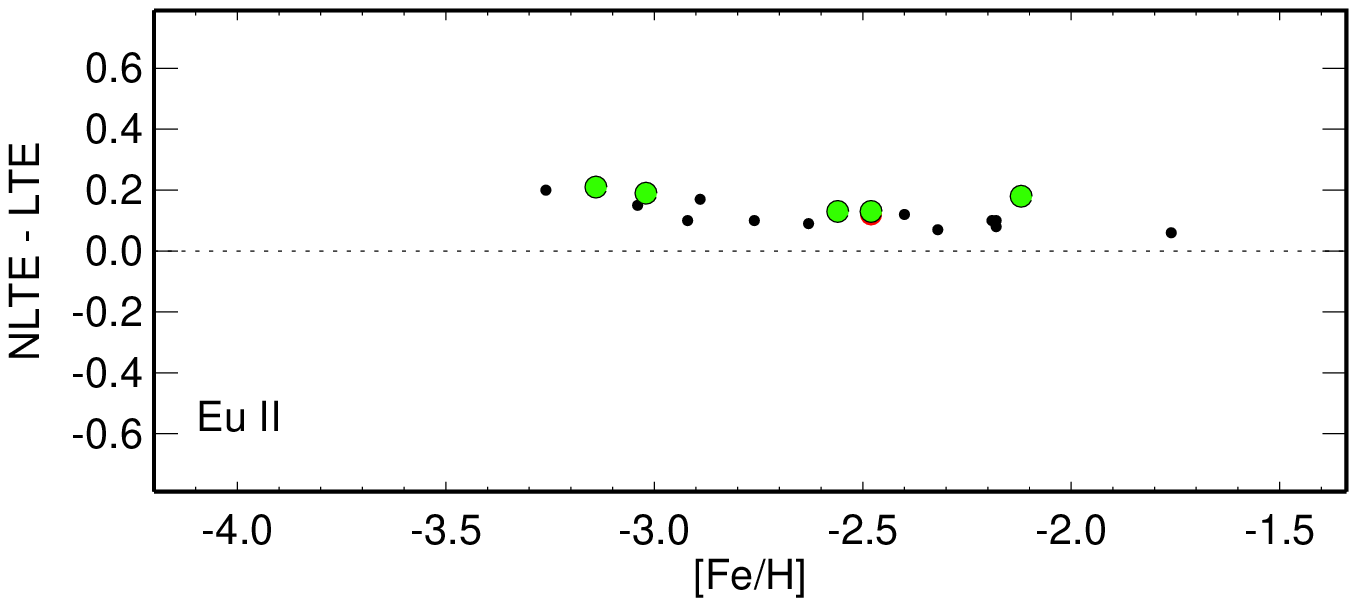}}
  \caption{\label{Fig:dnlte}  Differences between the NLTE and LTE abundance of different chemical species 
  in the Sculptor (red circles), Ursa Minor (green circles), Sextans (squares), Fornax (rhombi), Bo\"otes~I (triangles), UMa~II (inverted triangles), and Leo~IV (5 pointed star) dSph and the MW halo stars (small black circles). }
\end{figure*}

 Our NLTE calculations for \ion{Na}{i}, \ion{Mg}{i}, \ion{Al}{i},
  \ion{Si}{i}, \ion{Ca}{i-ii}, \ion{Ti}{i-ii}, \ion{Fe}{i-ii}, \ion{Sr}{ii}, and
  \ion{Ba}{ii} show that the departures from LTE are different for each species, and,
  in general, depend on the stellar atmosphere parameters and the element
  abundances.    
We stress that our procedure consistently accounts for both statistical
equilibrium and radiative transfer: the SE (NLTE) calculations were iterated by varying the element abundance until
agreement was reached between the resulting model spectra and the observed ones. 
  Figure\,\ref{Fig:dnlte} displays the differences between the
  average NLTE and LTE abundance, (NLTE~-~LTE), for the individual 
  stars. It is worth noting that this difference is set to the NLTE abundance correction,  $\Delta_{\rm NLTE}$, if we have one line, for example, for \ion{Al}{i} and \ion{Si}{i}.

We discuss different NLTE species by separating them depending on the dominant NLTE mechanism, but not in the order of their nuclear charge. The NLTE effects for \ion{Ti}{i} and \ion{Fe}{i} were described in our Paper~I.  For the reader convenience, we show the NLTE~-~LTE abundance differences for \ion{Ti}{i} and \ion{Fe}{i} also in this paper (Fig.\,\ref{Fig:dnlte}). For the range of stellar atmosphere parameters investigated in this study, all the NLTE
neutral species are minority ones. For example, even \ion{Si}{i}, which has the largest ionisation
energy, $E_{\rm ion}$ = 8.15~eV, contributes about 10~\%\ to the total Si
abundance in the line-formation layers of the model with $\Teff$/log~$g$/[Fe/H] = 4590/1.20/$-2.9$. The number density of these minority species easily deviates from
thermodynamic equilibrium, when the intensity of the ionising radiation deviates
from the Planck function. 

As discussed in the NLTE papers referenced in
Table~\ref{Tab:nlte}, the main NLTE mechanism for \ion{Mg}{i}, \ion{Al}{i}, and 
 \ion{Ca}{i} is the ultra-violet (UV)
over-ionisation. It results in the depletion of the atomic level populations and
weakened spectral lines compared to the LTE case and thus in positive NLTE
corrections.

\subsubsection{\ion{Al}{i}}

The largest positive NLTE corrections are computed for the resonance lines of \ion{Al}{i}. For example, $\Delta_{\rm NLTE}$(\ion{Al}{i} 3961\,\AA) reaches 0.66~dex for Scl6\_6\_402 (4890/1.78/$-3.66$). The NLTE effects for \ion{Al}{i} depend strongly on stellar $\Teff$, log~$g$, and metallicity (Al abundance), resulting in 
  much larger dispersion of the abundance differences between NLTE and LTE at fixed [Fe/H] compared with that for the other species, as shown in Fig.\,\ref{Fig:dnlte}. 
   For example, the MW stars HE1356-0622 (4945/2.0/$-3.45$) and HE1416-1032 (5000/2.0/$-3.23$) have  larger $\Delta_{\rm NLTE}$(\ion{Al}{i} 3961\,\AA), by about 0.3~dex, compared with that for the stars of similar metallicity, but lower effective temperature
BS16550-087 (4750/1.5/$-3.40$) and HE0039-4154 (4780/1.6/$-3.26$). 

\subsubsection{\ion{Ca}{i}}

There is a clear metallicity trend of the NLTE effects for \ion{Ca}{i}, with NLTE$-$LTE approaching 0.3~dex, when [Fe/H] is close to $-4$. A thorough discussion of the NLTE abundance corrections for an extended list of the \ion{Ca}{i}
lines can be found in \citet{Mashonkina_dnlte2016}. 

We encountered a problem with the \ion{Ca}{i} 4226\,\AA\ resonance line that
gives a lower abundance than the subordinate lines.
For example, for Scl031\_11 (4670/1.13/$-3.69$), the abundance difference between
\ion{Ca}{i} 4226\,\AA\ and three subordinate lines amounts to $\Delta\eps{} =
-0.65$ in the LTE calculations and becomes even more negative ($-0.87$~dex) in
NLTE.  \citet{mash_ca} have first reported  a similar case for the [Fe/H] $\simeq -2$ dwarf stars, although with a smaller abundance difference. This problem was 
highlighted by \citet{2012A&A...541A.143S} for their sample of VMP giants and dwarfs. Our calculations show  
that, in the VMP atmospheres, the total NLTE effect is smaller for the resonance line
than for the subordinate ones. Overall over-ionisation of \ion{Ca}{i} in deep atmospheric
layers leads to weakened subordinate lines and also line  wings of \ion{Ca}{i}
4226\,\AA. However, the core of \ion{Ca}{i} 4226\,\AA\ forms in the uppermost
atmospheric layers, where the departure coefficient of the upper level drops
rapidly due to photon escape from the line itself, resulting in dropping the line
source function below the Planck function and enhanced absorption in the line
core. This compensates in part or fully the NLTE effects in the line wings. 
For example, in the 4670/1.13/$-3.69$ model, \ion{Ca}{i} 4226\,\AA\ has equal
equivalent widths in NLTE and LTE. 

In consequence of the above considerations, we did not use \ion{Ca}{i}
4226\,\AA\ for the abundance determinations. For Scl07-50 ([Fe/H] = $-4.0$),
Scl6\_6\_402 ([Fe/H] = $-3.66$), and Scl11\_1\_4296 ([Fe/H] = $-3.7$), their Ca
abundance is derived from \ion{Ca}{ii} 3933\,\AA. It is worth noting,
$\Delta\eps{}$(\ion{Ca}{ii} 3933\,\AA\ - \ion{Ca}{i} 4226\,\AA) in these stars
amounts to 0.59~dex, 0.15~dex, and 0.27~dex, respectively. 

\subsubsection{\ion{Mg}{i}}

 As can be seen in Fig.\,\ref{Fig:dnlte}, NLTE$-$LTE is
  mostly positive for \ion{Mg}{i}. However, it is negative for Scl07-49 and
  Fnx05-42, where only the \ion{Mg}{i}b lines could be employed, and for the
  [Fe/H] $> -2.4$ stars, for which the used \ion{Mg}{i} 4703\,\AA\ and 5528\,\AA\ lines
  are strong ($W_{obs} > 100$~m\AA).
Strengthening the \ion{Mg}{i}b lines as compared to the LTE case was discussed
  in detail by \citet{mash_mg13} in her analysis of HD~122563. The case is the same as that of \ion{Ca}{i} 4226\,\AA\ reported in the previous subsection. We find that similar NLTE mechanisms act 
for \ion{Mg}{i} 4703\,\AA\ and 5528\,\AA, when they are strong.
Due to competing NLTE mechanisms, NLTE$-$LTE
for \ion{Mg}{i} is overall small, less than 0.12~dex in absolute value, except for 
Scl07-50 and Scl11\_1\_4296, for which all the \ion{Mg}{i} lines are weak and form in
deep atmospheric layers being subject to an over-ionisation of \ion{Mg}{i}.  

We note that, for \ion{Mg}{i}, NLTE leads to smaller line-to-line scatter compared
to the LTE case and cancels much of the scatter in [Mg/Fe] between stars of
    close metallicities. Hereafter, the sample standard deviation, $\sigma_{\eps{}} = \sqrt{\Sigma(\overline{x}-x_i)^2 / (N_l-1)}$, determines 
the dispersion in the
single line measurements around the mean.
For example, $\sigma_{\eps{}}$ = 0.06~dex in LTE and 0.03~dex in NLTE for HD~218857 (three lines of
\ion{Mg}{i}), and the abundance difference between HD~8724 ([Fe/H) = $-1.76$) and HD~218857
  ([Fe/H) = $-1.92$) amounts to $\Delta$[Mg/Fe] = 0.24~dex in LTE and 0.09~dex in NLTE. 

\subsubsection{\ion{Si}{i}}

Only small and mostly negative NLTE corrections are found for
the \ion{Si}{i} 3905 and 4102\,\AA\ lines, not exceeding
0.08~dex in absolute value. Indeed, with the accurate rate
coefficients from \citet{Belyaev2014_Si}, the inelastic collisions with the neutral
hydrogen atoms serve as efficient thermalising process and largely
cancel the over-ionisation effect for \ion{Si}{i} in line formation
layers.

\subsubsection{\ion{Na}{i}}\label{Sect:Na}

In contrast to the photoionisation-dominated minority species,
\ion{Na}{i} is subject to over-recombination because the photon suction
process prevails over the photoionisation that is inefficient for
\ion{Na}{i} due to small cross-sections of the ground state. The
over-recombination results in strengthening the \ion{Na}{i} 5889,
5895\,\AA\ lines and negative $\Delta_{\rm NLTE}$. The NLTE abundance
corrections vary between $-0.6$ and $-0.01$~dex. An exception is Scl031\_11
(4670/1.13/$-3.69$), for which the collision processes are inefficient
compared to the photoionisation. Hence, the \ion{Na}{i} lines are
weaker than in LTE and NLTE$-$LTE = +0.05~dex.

Only in HD~8724 ([Fe/H] = $-1.76$), could we measure not only the
\ion{Na}{i} resonance lines, but also the subordinate lines at
5682 and 5688\,\AA. The NLTE treatment of \ion{Na}{i}
substantially decreases the line-to-line scatter ($\sigma_{\eps{}}$ = 0.05~dex) compared with LTE, where $\sigma_{\eps{}}$ = 0.10~dex. A small abundance
difference of $-0.14$~dex between NLTE and LTE for HD~8724 is explained by including in the abundance mean 
the subordinate lines, for which $\Delta_{\rm NLTE}$ is
smaller, in absolute value, than that for the D$_1$, D$_2$
\ion{Na}{i} lines, by 0.1~dex, and also by cooler $\Teff$ of HD~8724 compared with that of  the [Fe/H] $\simeq -2$ stars.

Similarly to \ion{Al}{i}, \ion{Na}{i} shows a much larger scatter of
(NLTE--LTE) at fixed [Fe/H] than the other species. This reflects a
strong sensitivity of the departures from LTE for \ion{Na}{i} to stellar log~$g$, $\Teff$, and the Na abundance. For
example, at [Fe/H] $> -2.7$ some of the Ursa Minor (28104:
4275/0.65/$-2.12$), Sextans (11-04: 4380/0.57/$-2.60$), and Sculptor
(1019417: 4280/0.5/$-2.48$) stars do not follow the metallicity trend
 in Fig.\,\ref{Fig:dnlte} that is defined by the Milky
Way and Bo{\"o}tes~I population because of the lower density of their
atmospheres that weakens collisional coupling of the \ion{Na}{i}
high-excitation levels to the large continuum reservoir. For comparison, at [Fe/H] $\simeq -2.2$, HD~108317 and Boo-130 have log~$g$ = 2.96 and 1.4, respectively.
  

\subsubsection{\ion{Sr}{ii}, \ion{Ba}{ii}, and \ion{Eu}{ii}}

For the majority species such as \ion{Sr}{ii} and \ion{Ba}{ii}, NLTE may either
strengthen or weaken the line depending on the stellar parameter and
elemental abundance, as theoretically predicted by
\citet[][\ion{Ba}{ii}]{Mashonkina1999} and
\citet[][\ion{Sr}{ii}]{Mashonkina2001sr}.  In those MW halo giants, in which the Sr abundance follows the Fe one, the
\ion{Sr}{ii} 4077 and 4215\,\AA\ resonance lines are strong and the departures from LTE are small,
such that $\Delta_{\rm NLTE}$ varies between $-0.15$ and 0.08~dex. In the rest of the sample, the weaker the \ion{Sr}{ii} lines, the more positive NLTE correction is, up to
$\Delta_{\rm NLTE}$ = 0.35~dex.


A clear metallicity trend is
seen in Fig.\,\ref{Fig:dnlte} for barium, with negative NLTE$-$LTE for the [Fe/H] $> -2$ stars and
positive one for the majority of the VMP stars.
Since a change in the sign of the NLTE abundance correction for \ion{Ba}{ii} lines depends on the stellar parameters and the element abundance itself, some of the [Fe/H] $< -2$ stars have slightly negative differences NLTE$-$LTE.

As discussed by \citet{mash_eu}, NLTE leads to weakened lines of \ion{Eu}{ii} and positive NLTE abundance corrections. In our cool giant sample, $\Delta_{\rm NLTE}$ grows slowly towards lower metallicity, but never exceeds 0.2~dex.

\subsection{Nickel}\label{Sect:Ni}

Nickel is observed in lines of neutral atoms that are expected to be subject to
over-ionisation like other minority species, such as \ion{Al}{i}, \ion{Fe}{i},
etc. However, we cannot perform yet the NLTE calculations for \ion{Ni}{i}
because of the lack of a satisfactory model atom. Therefore, we assume
that the ratio of abundances derived from lines of \ion{Ni}{i} and \ion{Fe}{i} is nearly free of the NLTE
effects and use the LTE abundances in the [\ion{Ni}{i}/\ion{Fe}{i}] versus [Fe/H]
diagrams.  This assumption is supported by the fact that the ionisation energies of
\ion{Ni}{i} and \ion{Fe}{i} are very similar, 7.64~eV and 7.90~eV, respectively,
and these atoms have similarly complicated term structures.



\subsection{Outliers}\label{Sect:outlier}

In this section, we provide comments on a few stars, which reveal peculiar
abundances or any other outstanding feature.

{\it Sculptor ET0381.} This is an Fe-enhanced star, all measured chemical species
being deficient relative to Fe. The LTE abundances were discussed in detail in
\paperII. We only find here small departures from LTE for most species. Despite an
upward revision of [Fe/H], by 0.25~dex, leading to slightly changed [X/Fe] ratios,
we confirm the conclusions of \paperII. Except for [Ni/Fe], this star was not used in
the calculations of the average [X/Fe] ratios of the Sculptor dSph.

{\it Scl11\_1\_4296.} In this star, the $\alpha$-process elements Mg, Ca, and Ti
reveal a different behaviour with respect to Fe, namely, [Mg/Fe] is as low as that
of ET0381, however, [Ca/Fe] is higher compared with that of ET0381 and
close to solar value. Titanium is enhanced relative to Fe, similarly to the
remaining Sculptor dSph stars. This star was not used in the calculations of the
average [Mg/Fe] and [Ca/Fe] ratios of the Sculptor dSph.

{\it Sex24-72.} This is a carbon enhanced star. \paperI\ determined [C/Fe] = 0.4,
which applying the carbon correction from \citet{2014ApJ...797...21P} leads to an
initial [C/Fe] $\simeq$ 1. It reveals also high Na abundance, with [Na/Fe] = 0.85
(NLTE). Combined with low abundance of the neutron-capture elements ([Sr/Fe] =
$-0.43$, [Ba/Fe] = $-1.04$), Sex24-72 can be classified as a CEMP-no star.

{\it Boo-041.} We obtained higher iron abundance than that of
\citet{Gilmore2013}, by 0.42~dex, despite using common $\Teff$ = 4750~K and log~$g$ =
1.6. This cannot be explained by NLTE effects, because NLTE$-$LTE
= 0.06~dex for \ion{Fe}{i}. A difference of 0.25~dex in the average abundance from
the \ion{Fe}{i} lines appears already in LTE, as a consequence of lower microturbulence velocity, by 0.8\,\kms, in our study. In the LTE calculations
with $\Teff$ = 4750~K, log~$g$ = 1.6, and $\xi_t$ =
2.8\,\kms\ determined by \citet{Gilmore2013} and using 35 lines of
\ion{Fe}{i} with \Eexc\ $>$ 1.2~eV and $W_{obs} <$ 180~m\AA, we obtained a steep negative slope
of $-0.41$ for the $\eps{FeI}$ - log~$W_{obs}/\lambda$ plot. Besides, the
abundance difference $\eps{FeI}$ - $\eps{FeII}$ = $-0.22$ was uncomfortably large.
We established $\xi_t$ = 2.0\,\kms\ by minimising the slope of the \ion{Fe}{i}-based NLTE
abundance trend with $W_{obs}$. This also leads to consistent NLTE abundances from the two
ionisation stages of iron within 0.11~dex. This star reveals extremely high
abundance of Ti, with [Ti/Fe] = 0.80, but low abundance of Ni, with [Ni~I/Fe~I] =
$-0.52$ (LTE). Boo-041 was not used in the calculations of the average [X/Fe]
ratios. At [Fe/H] = $-1.54$, the Fe abundance of Boo-041 might have received the
products of the type Ia supernova (SN~Ia) nucleosynthesis.

\subsection{Influence of uncertainties in stellar atmosphere parameters}

Changes in the element abundances caused by a variation in $\Teff$, log~$g$, and $\xi_t$ were evaluated for a part of our stellar sample in the earlier LTE analyses of \citet[][Table~4]{2010A&A...524A..58T} and \citet[][Tables~5 and 6]{2015A&A...583A..67J} and the NLTE analyses of \citet[][Table~6]{HE2327} and \citet[][Sect.~4.2.4]{dsph_parameters}. When varying $\Teff$, a differential NLTE effect on the derived abundance is the largest for \ion{Al}{i} 3961\,\AA, however, it does not exceed 0.03 and 0.05~dex for $\Delta\Teff$ = +100~K around $\Teff$ = 4500 and 5000~K, respectively. Since the sample stars have close together temperatures and surface gravities, they have also close together abundance errors due to uncertainties in atmospheric parameters. To summarise, a change of +100~K in $\Teff$ produces higher abundances from lines of the minority species, such as \ion{Mg}{i} and \ion{Al}{i}, by 0.10-0.15~dex, and
has a minor effect ($\le$ 0.02~dex) on the abundances from lines of the majority species, such as \ion{Ti}{ii} and \ion{Sr}{ii}. In contrast, a change of +0.1~dex in log~$g$ has a minor effect ($\le$ 0.01~dex) on the minority species and shifts abundances of the majority species  by up to +0.05~dex. A change of +0.2\,\kms\ in $\xi_t$ produces lower abundances, by 0.1~dex for the EW $\simeq$ 120\,m\AA\ lines and by 0.05~dex, if EW $\simeq$ 75\,m\AA.

\section{Abundance trends and galaxy comparisons}
\label{Sect:dsph_halo}

\begin{figure}  
\begin{center}
 \hspace{-6mm}  \resizebox{95mm}{!}{\includegraphics{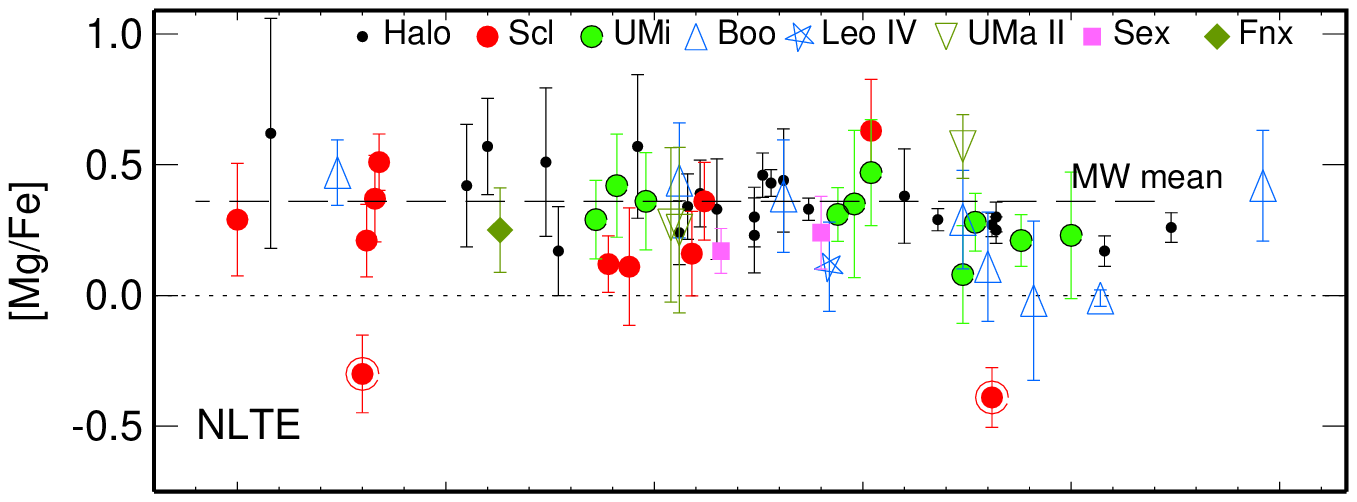}}

  \vspace{-13mm}
 \hspace{-6mm}   \resizebox{95mm}{!}{\includegraphics{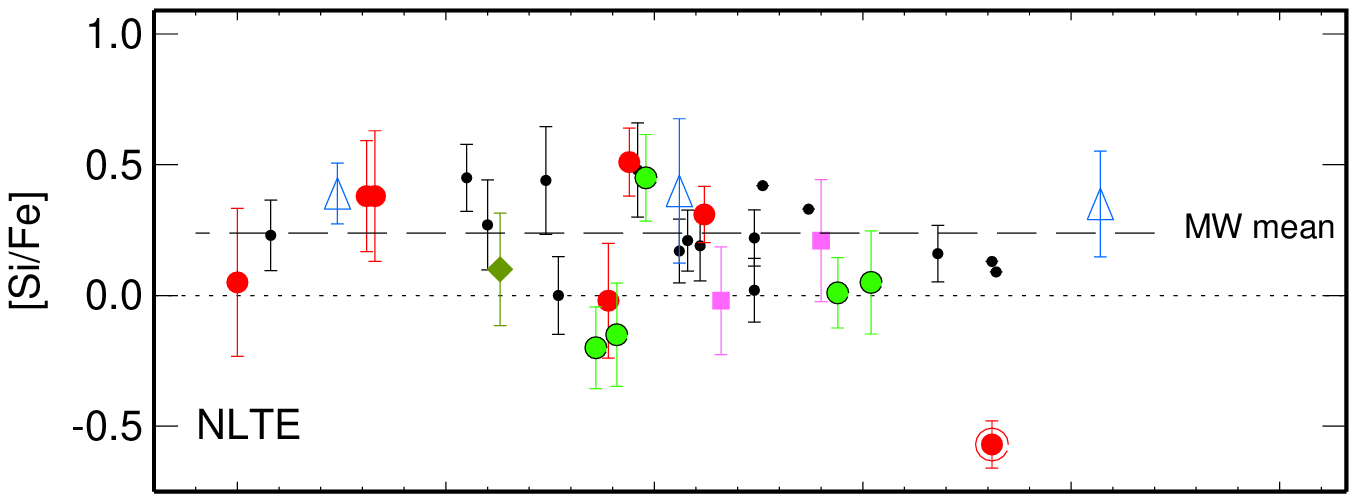}}  

  \vspace{-13mm}
 \hspace{-6mm}   \resizebox{95mm}{!}{\includegraphics{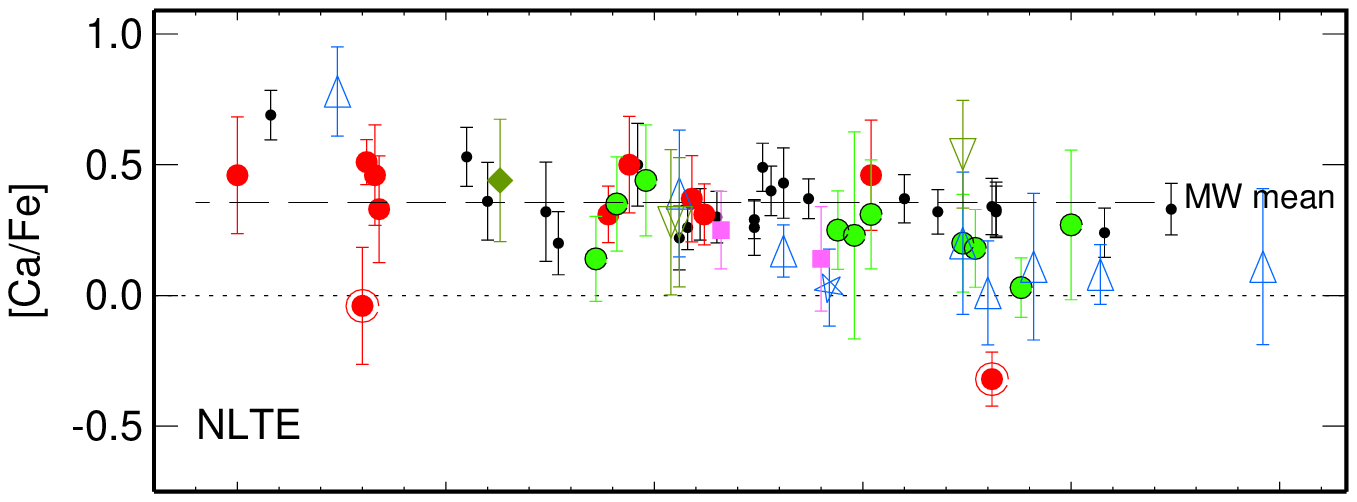}}

  \vspace{-13mm}
 \hspace{-6mm}   \resizebox{95mm}{!}{\includegraphics{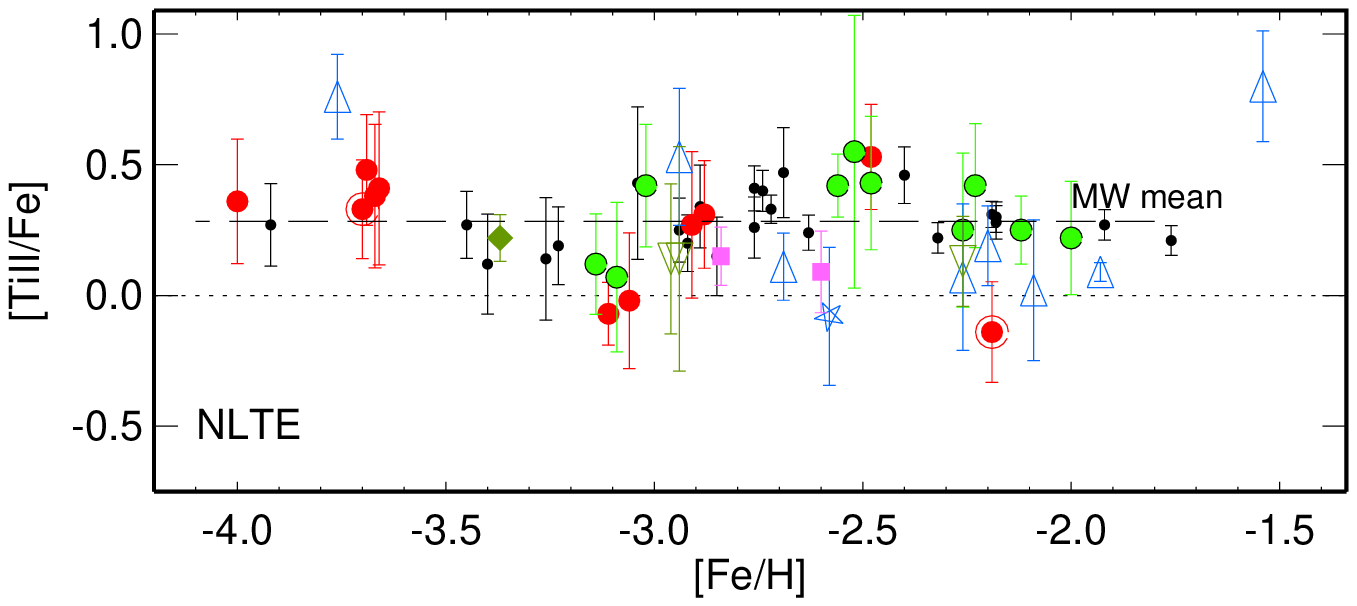}}

  \caption{\label{Fig:alpha} Stellar element-to-iron NLTE abundance ratios for the $\alpha$-process elements Mg, Si, Ca, and Ti in the
 Sculptor (red circles), Ursa Minor (green circles), Sextans (squares), Fornax (rhombi), Bo\"otes~I (triangles), UMa~II (inverted triangles), and Leo~IV (5 pointed star)
dSphs and the MW halo. Two Mg-poor stars in Sculptor, 11\_1\_4296 and ET0381, are plotted by a red circle inside larger size open circle. 
The error bars were computed as $\sigma_{\rm X/Y} = \sqrt{\sigma_{\rm X}^2 + \sigma_{\rm Y}^2}$. The errors of [Fe/H] do not exceed 0.15~dex for most (51 of 59) stars.
In each panel, the dashed line indicates the mean for the MW halo. 
}
\end{center}
\end{figure}

\begin{figure}  
\hspace{-5mm}  \resizebox{95mm}{!}{\includegraphics{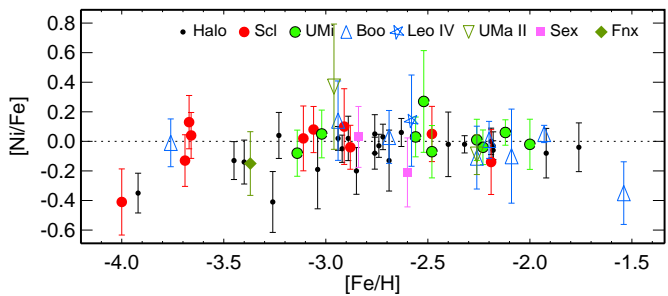}}
  \caption{\label{Fig:ni_zn} Stellar [Ni~I/Fe~I] abundance ratios. Symbols as in Fig.~\ref{Fig:alpha}. }
\end{figure}

Table~\ref{Tab:AbundanceSummary} presents the elemental
 LTE and NLTE abundances together with their
statistical errors and the number of lines used. For
consistency with our previous studies, the solar photosphere
abundances are those of \citet{AG1989} at the exception of Ti and Fe,
for which we consider $\eps{Ti,met}$ = 4.93 \citep[][meteoritic
  abundance]{Lodders2009} and $\eps{Fe,\odot}$ = 7.50
\citep{1998SSRv...85..161G}. The metallicities, [Fe/H], as well as the
abundance ratios relative to iron, [X/Fe], are based on the
\ion{Fe}{ii} lines. Our results are displayed in
Figs.\,\ref{Fig:alpha}-\ref{Fig:srbamg_bah}.

\begin{table} 
 \caption{\label{Tab:mean} Average NLTE abundance ratios for different stellar populations, with $\sigma_{\rm X/Fe}$ indicated in parentheses.}
 \centering
 \begin{tabular}{lrcrrrrrr}
\hline\hline \noalign{\smallskip}
 [X/Fe]  & \multicolumn{1}{c}{Sculptor}   & Ursa Minor & MW halo \\
\noalign{\smallskip} \hline \noalign{\smallskip}
 \ [Mg/Fe] & 0.31 (0.19) & 0.30 (0.11)   & 0.36 (0.13) \\
 \ [Si/Fe] & 0.25 (0.21) & 0.03 (0.26)   & 0.24 (0.15) \\
 \ [Ca/Fe] & 0.41 (0.08) & 0.24 (0.11)   & 0.36 (0.11) \\
 \ [Ti/Fe] & 0.30 (0.19) & 0.32 (0.16)   & 0.28 (0.10) \\
\noalign{\smallskip}\hline \noalign{\smallskip}
\end{tabular}
\end{table}

\subsection{$\alpha$-process elements}\label{Sect:MW}

For all $\alpha$-process elements, we refer to their neutral species, except for Ti for which the abundance is based on lines of
\ion{Ti}{ii}. Indeed the \ion{Ti}{ii} lines are more numerous and stronger than
the \ion{Ti}{i} ones. They are also less affected by any departure from LTE. We
note though that, in most of our sample stars, the abundances derived from the
lines of \ion{Ti}{i} and \ion{Ti}{ii} are consistent, as shown in Paper~I.

Before going further, it is probably worth commenting on the consequence of
choosing \ion{Fe}{ii} or \ion{Fe}{i} as a metallicity indicator on the [X/Fe] versus
[Fe/H] diagrams. Indeed, the vast majority of the published LTE analyses are
provided in function of [\ion{Fe}{i}/H]. This is a consequence of the fact that
the number of \ion{Fe}{i} lines is classically much larger in the observed
wavelength range. 
As shown in Fig.\,\ref{Fig:dnlte}, the LTE treatment underestimates the iron abundance, if it is based on the \ion{Fe}{i} lines, and NLTE$-$LTE (\ion{Fe}{i}) rises towards lower metallicity. For \ion{Ca}{i}, NLTE$-$LTE follows a similar trend and
magnitude as \ion{Fe}{i}, hence NLTE leaves their ratios nearly
unchanged compared with LTE. Conversely, because the NLTE corrections for \ion{Mg}{i} and
\ion{Si}{i} are minor over the [$-4$, $-1$] metallicity
range, [\ion{Mg}{i}/\ion{Fe}{i}] and [\ion{Si}{i}/\ion{Fe}{i}] are
shifted downward in NLTE as compared to LTE. 

Now, choosing \ion{Fe}{ii} as a metallicity indicator, 
[\ion{Mg}{i}/\ion{Fe}{ii}], [\ion{Si}{i}/\ion{Fe}{ii}], and
[\ion{Ti}{ii}/\ion{Fe}{ii}] change only a little from LTE to NLTE,
while [\ion{Ca}{i}/\ion{Fe}{ii}] moves upward, because of the positive NLTE
corrections for \ion{Ca}{i}. 

Because, in the NLTE treatment, we have obtained consistent \ion{Fe}{i}- and \ion{Fe}{ii}-based  abundances (Paper~I), our NLTE abundance ratios [X/Fe] do not depend on the choice of the metallicity tracer.  


A remarkable gain of using the NLTE abundances based on a homogeneous set of atmospheric parameters is the reduction,
compared to a simple compilation of the literature data \citep[see Figs.\,8 and 9 in][]{2015A&A...583A..67J}, of the spread in
abundance ratios at given metallicity within each galaxy and from one to the
other. This effect is particularly dramatic for Si for which the dispersion goes
from a $+1$~dex down to $\sim$0.2~dex once the Si and Fe abundances
have been homogeneously revised.


Figure\,\ref{Fig:alpha} shows that, at [Fe/H] $\le -2.5$, all galaxies scatter around the mean of the
Milky Way halo stars, [$\alpha$/Fe] $\simeq$ 0.3. Table\,\ref{Tab:mean} summarizes the mean elemental ratios and their dispersions 
for the three galaxies with a sufficient number of stars: Sculptor, Ursa Minor, and the MW halo. The outliers of
Sect.~\ref{Sect:outlier} are not considered in these calculations. These numbers
quantify the small star-to-star scatter ($\sim$ 0.2~dex) and show that all $\alpha$-elements have similar [X/Fe] means. The only apparent exception to this rule is
the solar value of [Si/Fe] in Ursa Minor. For the JI19 star, which is the
only one in this galaxy with two accessible \ion{Si}{i} lines, [Si/Fe] =
0.45 ($\sigma_{\eps{}}$ = 0.07~dex) is close to [X/Fe] for the other $\alpha$-elements. The four other stars in Ursa
Minor have only one \ion{Si}{i} line at 4102\,\AA, which is located
in a noisy region and which leads to a solar or subsolar [Si/Fe]
ratio. It is worth noting that our results for the MW giants are fully consistent with the
NLTE abundances derived for the MW halo dwarfs by \citet{lick_paperII}.

 Previous LTE abundance analyses conducted at similar high
spectral resolution and signal-to-noise ratios, by
 \citet{2010A&A...524A..58T,Starkenburg2013,2015A&A...583A..67J}, and
 \citet{2015ApJ...802...93S}, deduced a common conclusion that the Sculptor,
 Sextans, and Fornax dSphs are $\alpha$-enhanced in the VMP regime, at similar
 level to the MW halo. On the other hand, lower [Ca/Fe] and [Ti/Fe] than [Mg/Fe] ratios were
 reported by \citet{2004AJ....128.1177V} and \citet{2010ApJ...719..931C} for the
 VMP stars in the Sextans and Ursa Minor dSphs. 
 In part, these discrepancies can be caused by applying the LTE assumption. In case of using lines of \ion{Ti}{i} and \ion{Fe}{i}, LTE underestimates [Ti/Fe] because of larger departures from LTE for the \ion{Ti}{i} than for the \ion{Fe}{i} lines. In contrast, [\ion{Mg}{i}/\ion{Fe}{i}] is overestimated in the LTE analysis, as discussed in the beginning of this section.
 Our homogeneous NLTE
 analysis removes discrepancies in [X/Fe] between different $\alpha$-elements
 and between the classical dSphs and the MW halo. 

We have three ultra-faint galaxies ($L \le$ 10$^5$ $L_{\odot}$) in our sample:
Ursa Major~II, Leo~IV, and Bo\"otes~I. Interestingly, they present different
features. In Ursa Major~II, the three stars with very different metallicities, between
[Fe/H] $\simeq -3$ and $-2.3$, are $\alpha$-enhanced. In contrast, Leo~IV and
Bo\"otes~I, which are the brightest of the Local Group UFDs
\citep{2012AJ....144....4M}, reveal a close-to-solar $\alpha$/Fe ratio at
      [Fe/H]$\gtrsim -2.5$. \citet{Gilmore2013} had reported on a hint of a decline
      in [$\alpha$/Fe] but not yet ``formally significant''. The existence of a low [$\alpha$/Fe]
      population is now put on a firm ground, with consistent evidence from the
      three elements, for which data are available, Mg, Ca, and Ti. We 
      commented earlier that Boo-041 cannot be accounted for the bulk of
      Bo\"otes~I population. The kinematics and morphology of Bo\"otes~I point
      towards a complex system \citep{2011ApJ...736..146K, 2016MNRAS.461.3702R},
      but as far as its chemical evolution is concerned, Bo\"otes~I, and
      potentially Leo~IV as well, seem to simply push back the frontier, at
      which galaxies can reprocess the SNeIa ejecta, with a knee at lower
      metallicity than their more massive counterparts, as expected from a
      classical chemical evolution. So far the galaxies, in which a knee in
      [$\alpha$/Fe] has been found, had star formation histories of a few Gyrs
      long, while those of Bo\"otes~I and Leo~IV are shorter
      \citep{2012ApJ...744...96O}. However, the nature and timescale of the type
      Ia supernovae are still under investigation
      \citep{2009ApJ...707.1466K,2013ARA&A..51..457N}, with the evidence for metal-poor
      environments having their own specific features
      \citep{2011MNRAS.412.2735T}. Given that star formation proceeds in series
      of very short timescale bursts \citep[$\sim$0.1~Gyr,][]{Revaz:2012cw}, it
      is very possible that Bo\"otes~I had enough time to form stars after its first
      SN~Ia explosions, provided the latter occur on short timescales. 
      The chemical evolution models by \citet{2015MNRAS.446.4220R} allow some SNeIa
      to be exploded in Bo\"otes~I while it was still forming stars.  An
      alternative possibility is that low mass SNeII have particularly
      contributed to the chemical evolution of the galaxy \citep{2015ApJ...799L..21W}.
            


\subsection{Nickel}

As discussed in Sect.\,\ref{Sect:Ni}, we did not treat Ni in NLTE. Nevertheless,
we think that the [\ion{Ni}{i}/\ion{Fe}{i}] versus [Fe/H] trend is relevant to this work. Figure \ref{Fig:ni_zn}
    shows that the yields of Ni and Fe have a constant ratio and the
    correlation is reasonably tight. It is, in fact, tighter than with any other
    element produced by SNeII. Both dSph and MW halo stars are distributed
    around [Ni/Fe] = 0. We do not find in either population the high [Ni/Fe] stars
as in \citet{2009AJ....137..272R}. Given the large range of galaxy masses that we
are sampling, this is quite remarkable. The only trend with metallicity is the
increased dispersion at [Fe/H] $\le -3$, that is, 0.15~dex instead of 0.08~dex above [Fe/H] = $-3$,
which could well be explained by increasing the abundance uncertainties due to the weakening the \ion{Ni}{i} lines.


\begin{figure*}  
\hspace{-3mm}    \resizebox{95mm}{!}{\includegraphics{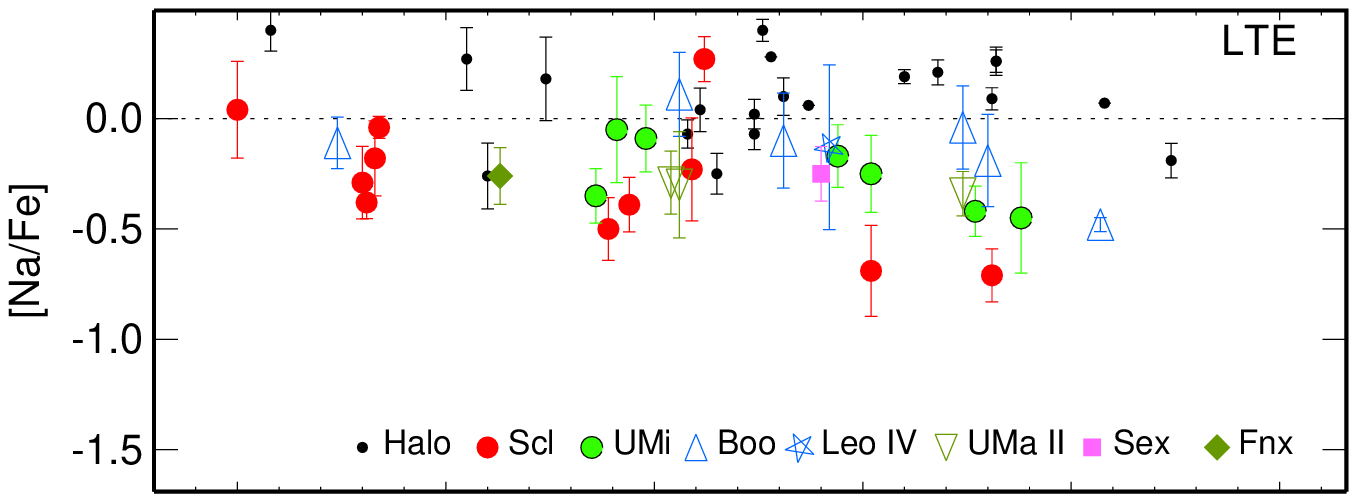}} 
\hspace{-3mm}    \resizebox{95mm}{!}{\includegraphics{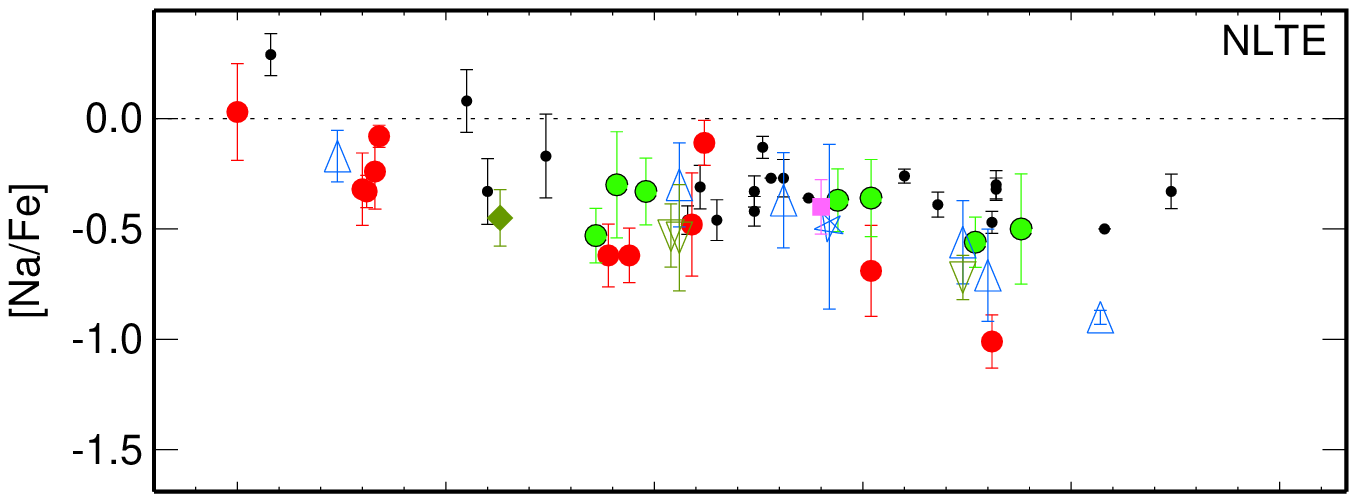}} 

\vspace{-13mm}
\hspace{-3mm}    \resizebox{95mm}{!}{\includegraphics{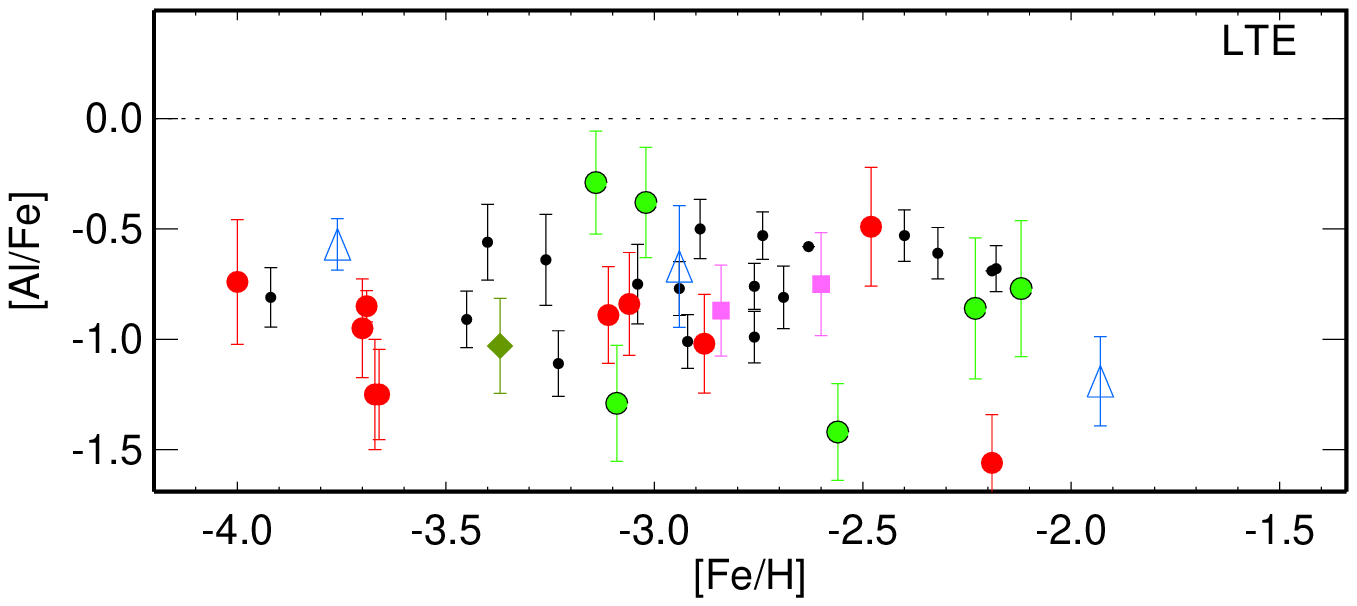}}
\hspace{-3mm}    \resizebox{95mm}{!}{\includegraphics{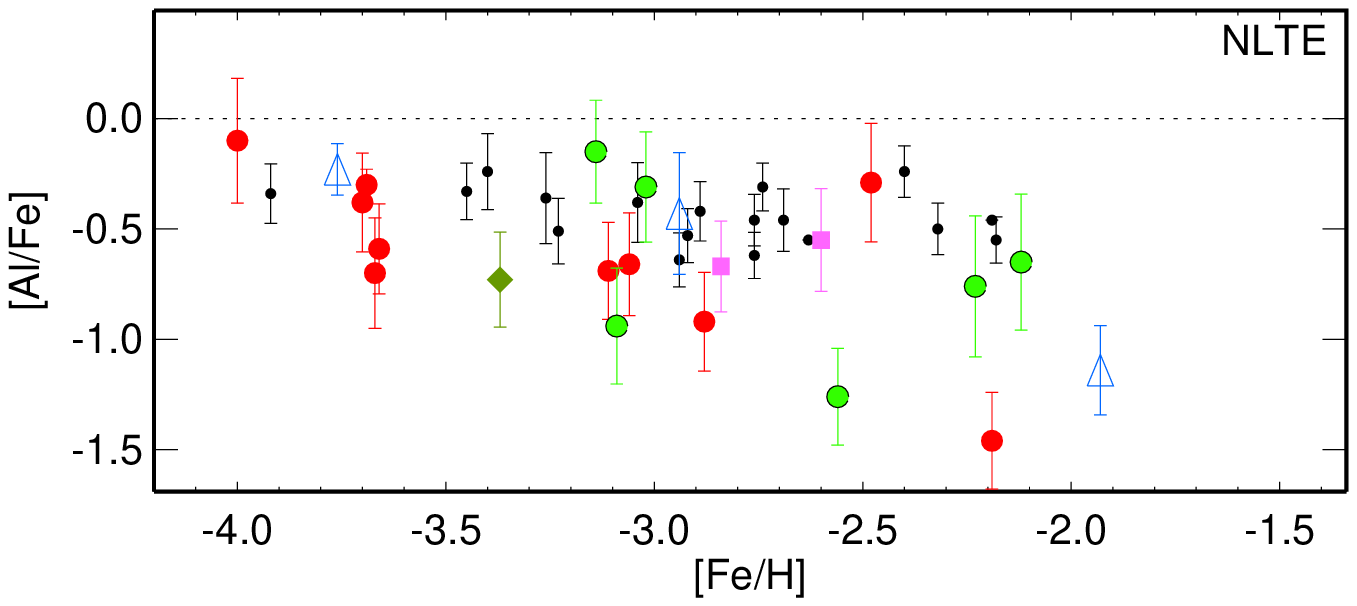}}
    \caption{\label{Fig:oddfe} Stellar [Na/Fe] and [Al/Fe] LTE (left column) and NLTE (right column) abundance ratios. For Sex~24-72 its LTE and NLTE ratios of [Na/Fe] = 0.99 and 0.85, respectively, are not displayed. Symbols as in Fig.~\ref{Fig:dnlte}.}
\end{figure*}

\begin{figure*}  
\hspace{-3mm}  \resizebox{95mm}{!}{\includegraphics{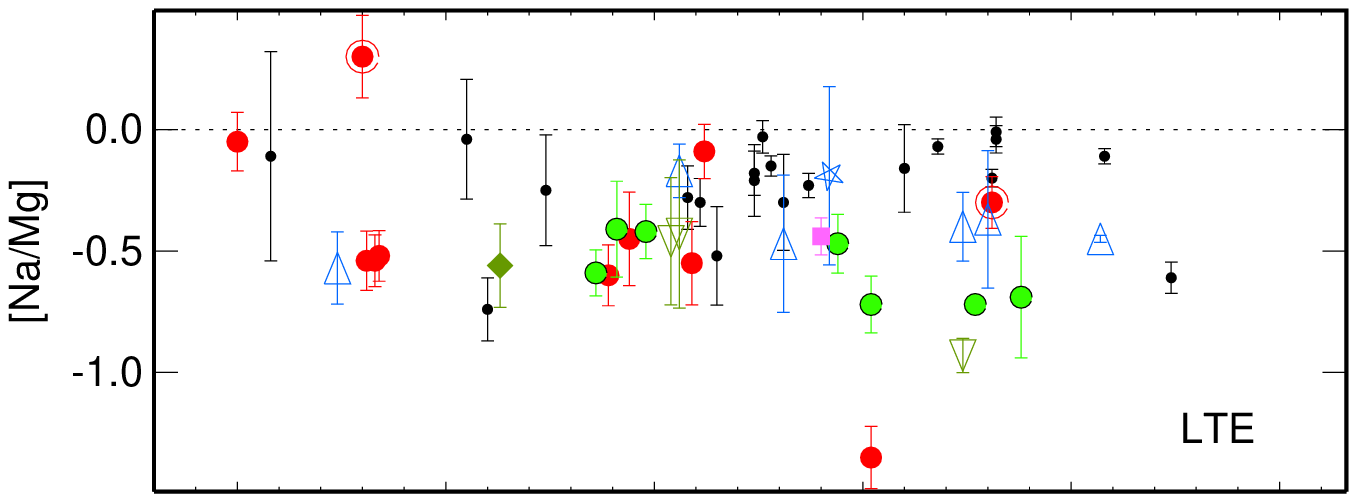}}
\hspace{-3mm}  \resizebox{95mm}{!}{\includegraphics{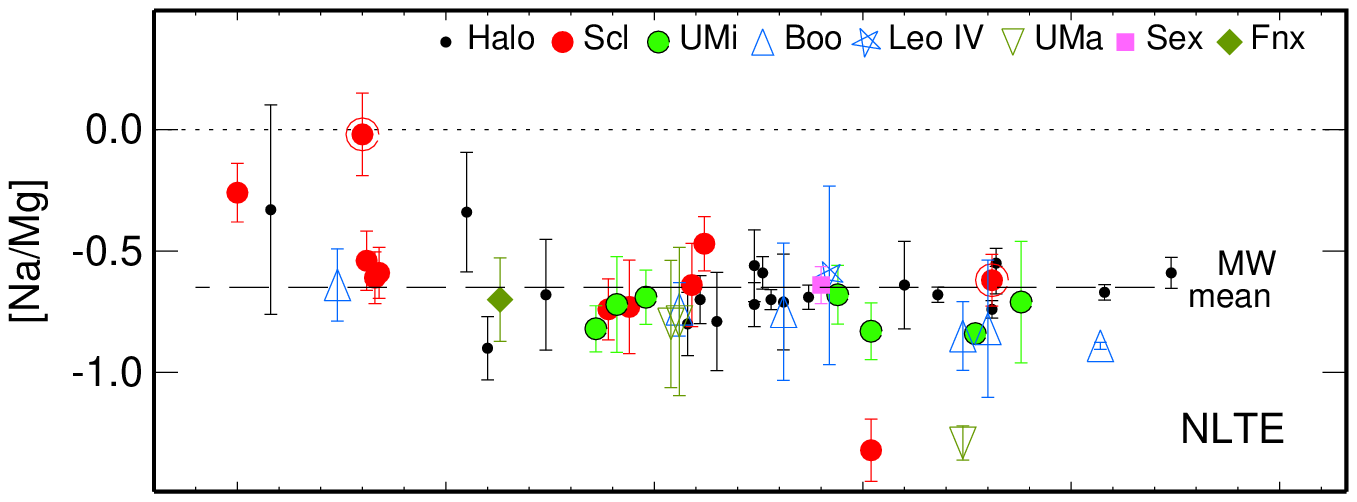}}

\vspace{-13mm}
\hspace{-3mm}    \resizebox{95mm}{!}{\includegraphics{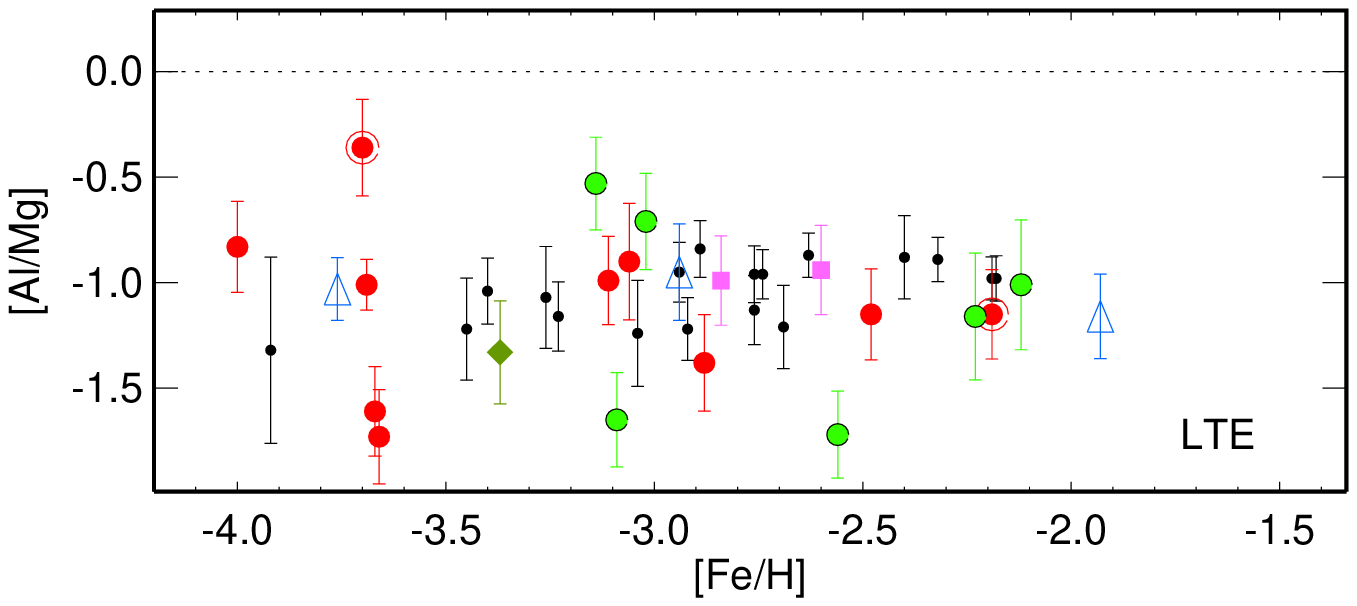}}
\hspace{-3mm}    \resizebox{95mm}{!}{\includegraphics{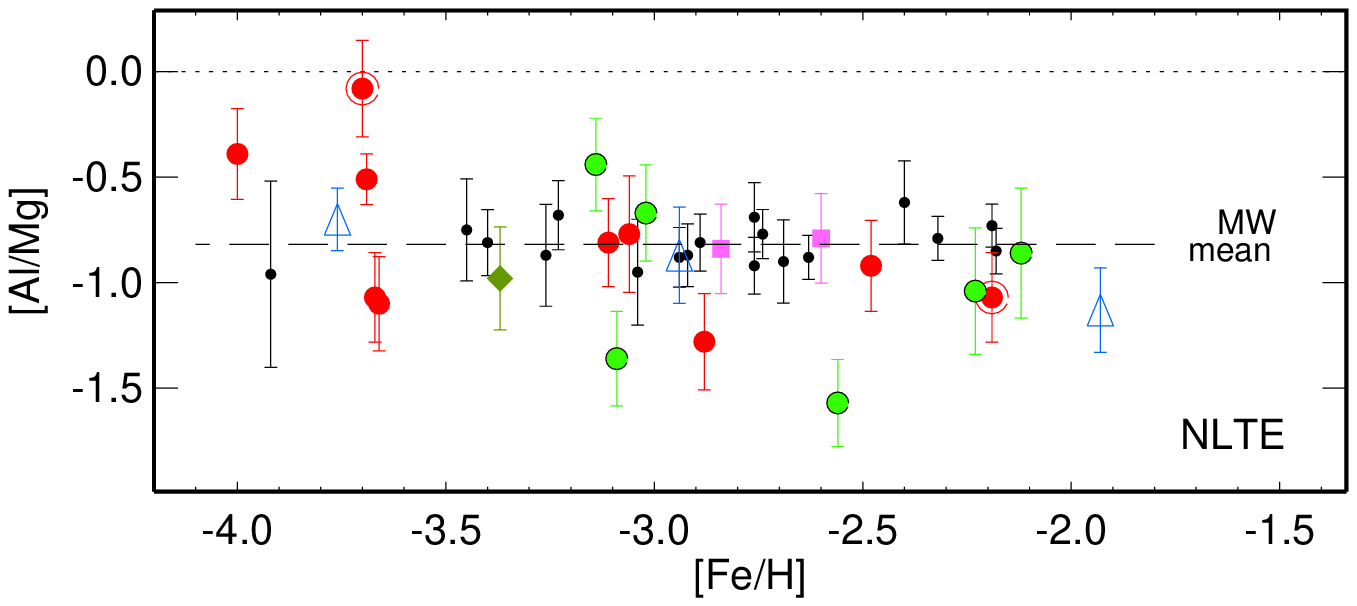}}  
  
    \caption{\label{Fig:odd} Stellar [Na/Mg] and [Al/Mg] LTE (left column) and NLTE (right column) abundance ratios. For Sex~24-72, its NLTE ratio [Na/Mg] = 0.68
is not displayed. Symbols as in Fig.~\ref{Fig:alpha}. }
\end{figure*}

\subsection{Odd-$Z$ elements: Na and Al}

The impact of NLTE is particularly large on the elemental ratios involving Na and Al,
as a consequence of large NLTE abundance corrections for lines of \ion{Na}{i} and \ion{Al}{i} and their extreme sensitivity to each of the stellar parameters. 
The star-to-star scatter in the [Na/Fe] (Fig.\,\ref{Fig:oddfe}) and [Na/Mg] (Fig.\,\ref{Fig:odd}) NLTE abundance ratios is much smaller than in the LTE ones. For [Al/Fe] and [Al/Mg], the scatter is larger for the dSph than the MW stars, most probably, due to lower S/N ratio of the observed blue spectra and the uncertainty in analysis of \ion{Al}{i} 3961\,\AA.

For [Na/Fe], [Na/Mg], and [Al/Mg], the Milky Way and dSphs reveal indistinguishable trends with metallicity suggesting that the nucleosynthesis processes for Na and Al (carbon burning
process) are identical in all systems, independent of their mass. This is in contrast to LTE, where, for the MW stars, we obtain systematically higher [Na/Fe] and [Na/Mg] ratios than for the dSph stars, by about 0.4 and 0.3~dex, respectively.
It seems that the early production of sodium occurred in a way similar to that for the primary
elements. This is confirmed by the constant [Na/Mg] $\simeq -0.6$  and
[Al/Mg] $\simeq -0.8$ ratios in Fig.\,\ref{Fig:odd}.
  The only exception to this rule is Scl11\_1\_4296, whose high [odd-$Z$/Mg] ratios
  are only due to its depletion in Mg. The ET0381 star does not stand out in [odd-$Z$/Mg],
  meaning that its Al, Na, and Mg were produced in similar relative amounts, as these elements in the
  other stars.

  We confirm the existence of a plateau in [Na/Fe] at [Fe/H] $\le -2.0$ showed in
\citet{2007A&A...464.1081A}. These authors reported a mean value of
      [Na/Fe] = $-0.2$, while we find [Na/Fe] = $-0.4$.
The difference between the two studies arises, probably, from
      the fact that \citet{2007A&A...464.1081A} used the iron LTE abundances and, hence,
      underestimated [Fe/H] values.
Just as us here, \citet{2007A&A...464.1081A} showed a hint of a rise in [Na/Fe]
below [Fe/H] = $-3.5$. As discussed in Paper~I, the NLTE ionisation balance between
\ion{Fe}{i} and \ion{Fe}{ii} is not fully satisfied for the Sculptor stars in
this region. Hence we might actually underestimate [Fe/H] by a bit by referring
to \ion{Fe}{ii}. The Milky Way star at [Fe/H] = $-3.92$, HE1357-0123, with more than 0.2~dex difference between its \ion{Fe}{i}- and \ion{Fe}{ii}-based abundances, also follows the [Na/Fe] rising trend, calling for further investigation.

Going from LTE to NLTE changes dramatically the picture of early enrichment of
the MW and the dSphs in Na and Al. In NLTE, sodium is found to be slightly
overabundant relative to aluminum, with the mean of [Na/Al] $\sim$ 0.2~dex,
while substantially larger ratio of [Na/Al] $\sim$ 0.8 is obtained in LTE.

For clarity, we did not include a carbon-enhanced star Sex24-72 in the [Na/Fe] panel of
Fig.\,\ref{Fig:oddfe}. This is one of the very rare CEMP-no stars found in a
classical dSph \citep{2015A&A...574A.129S}. \citet{2010A&A...524A..58T} have
measured $^{12}$C/$^{13}$C = 6 in this star, a clear indication of internal mixing \citep[see, for example,][]{2006A&A...455..291S}. We
determine [Na/Fe] = 0.85, [Na/Mg] = 0.68, and [Na/Al] = 1.52, hence, a clear
overabundance of sodium. Unlike the mixed stars in \citet{2006A&A...455..291S},
 the [Al/Mg] ratio of Sex24-72 does not stand out from our other sample stars, and
this star is not particularly Mg poor. Therefore, the abundance pattern of Sex24-72 was not affected by the Mg-Al cycle.
We conclude that the Na overabundance of Sex24-72 is a consequence of the extra-mixing between the
atmosphere and the H-burning shell, which was deep enough to bring the products
of the Ne-Na cycle to the surface.

\begin{figure*}  
\hspace{-3mm}   \resizebox{95mm}{!}{\includegraphics{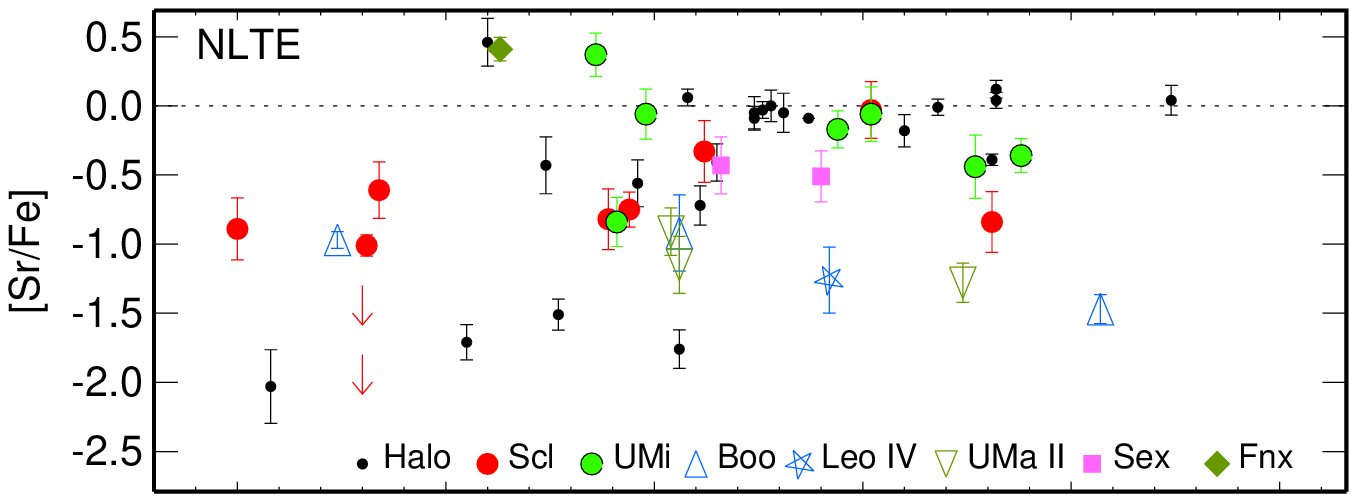}}
\hspace{-3mm}   \resizebox{95mm}{!}{\includegraphics{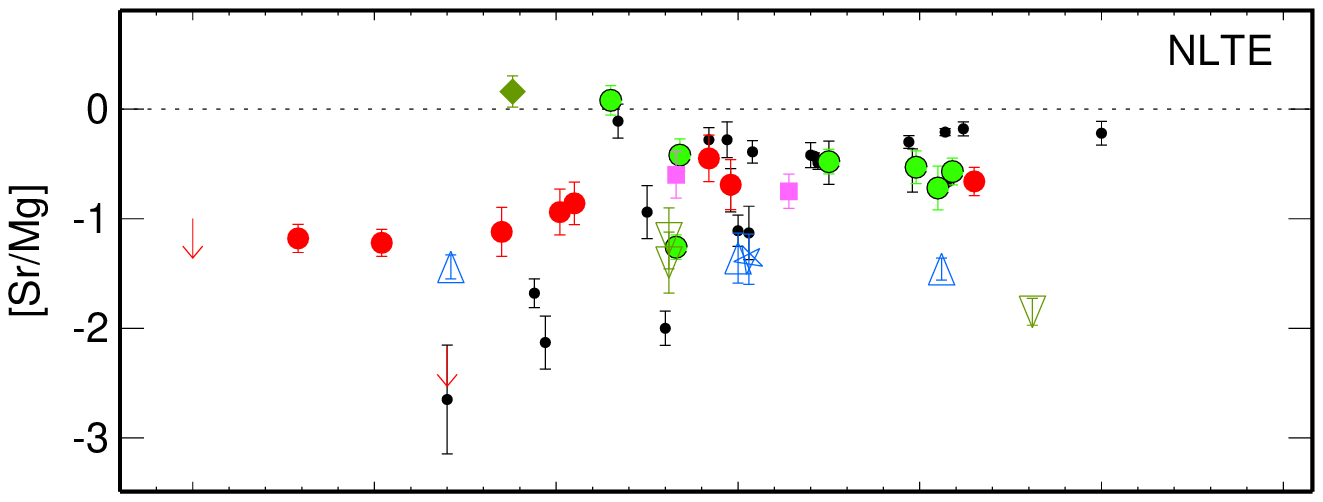}}

  \vspace{-13mm}
\hspace{-3mm}   \resizebox{95mm}{!}{\includegraphics{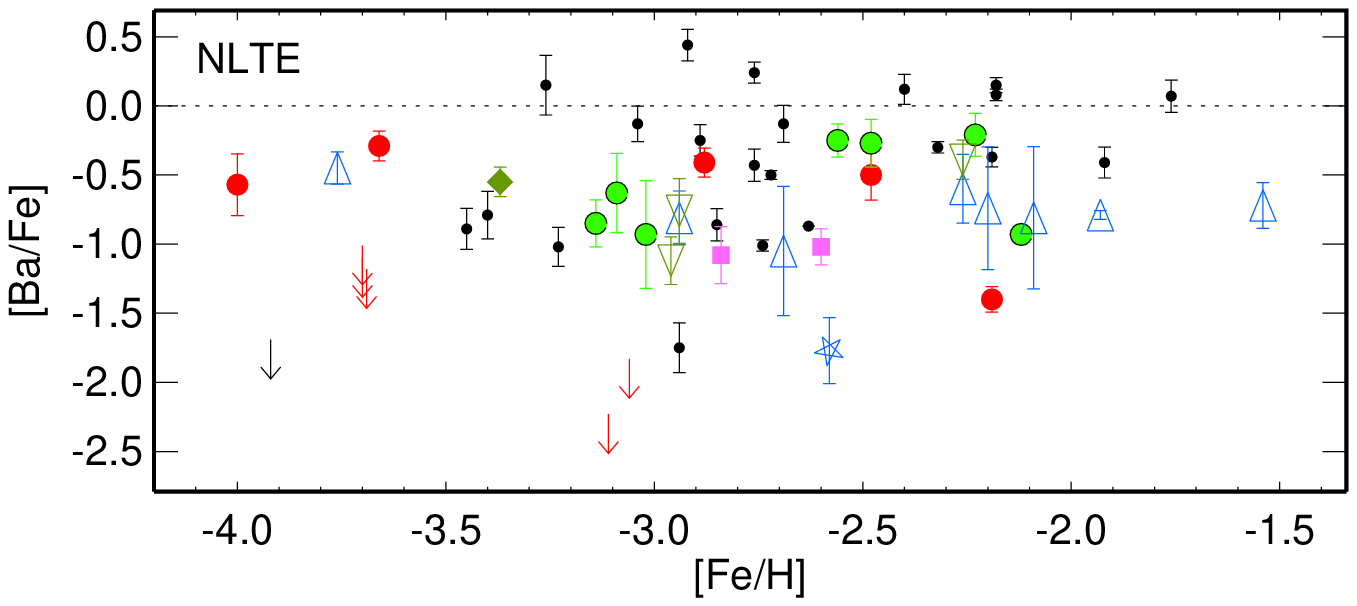}}
\hspace{-3mm}   \resizebox{95mm}{!}{\includegraphics{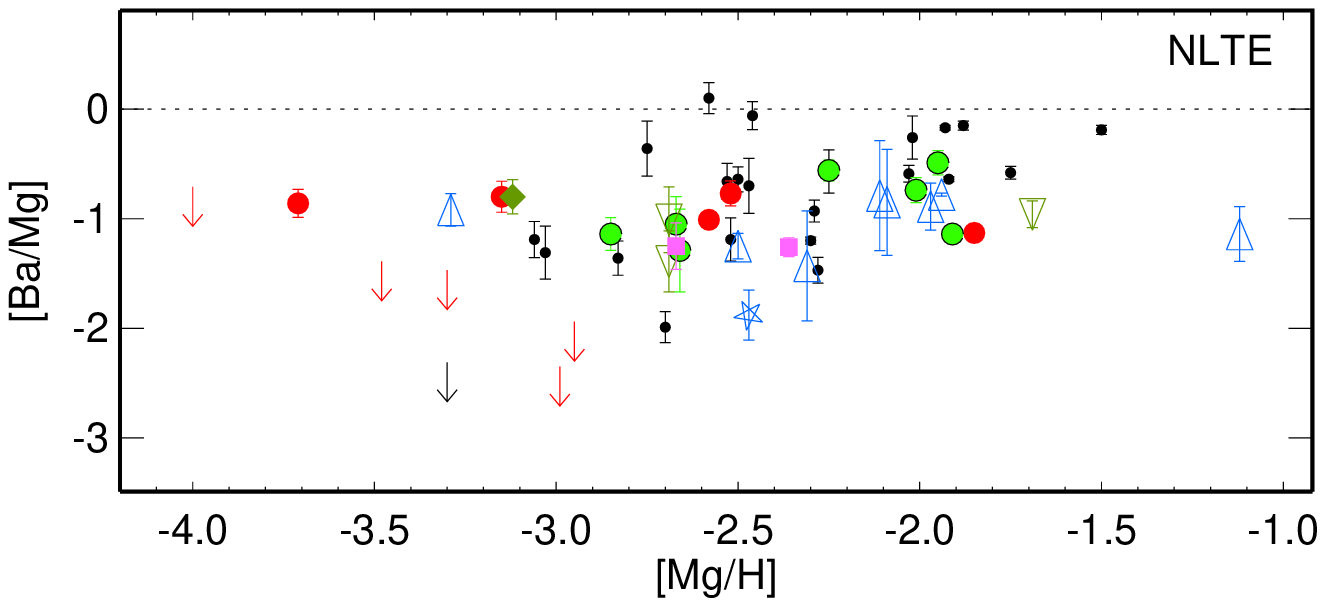}}
  \caption{\label{Fig:sr_ba} Left column: [Sr/Fe] and [Ba/Fe] NLTE abundance ratios as a function of metallicity for our sample stars. Right column: [Sr/Mg] and [Ba/Mg] NLTE abundance ratios as a function of Mg abundance. Symbols and colours are as in Fig.~\ref{Fig:alpha}. Upper limits in the Ba and Sr abundances are indicated with arrows.}
\end{figure*}

\subsection{Neutron-capture elements: Sr and Ba}

Despite many theoretical and observational studies of the neutron-capture elements in the long-lived stars in our Galaxy, 
there are not yet clear answers to the following questions: what is (are) the astrophysical site(s) of the rapid (r) process of neutron-capture nuclear reactions, what types of nuclear reactions produced the light trans-iron elements, Sr-Zr, in the early Universe and at what astrophysical site(s), and did the light and heavy (beyond Ba) elements originate from a common astrophysical site. 

It is beyond the scope of this work to provide any definitive answers. However,
the homogeneity of our abundance analysis and our compilation of galaxies with very different
evolutionary paths do provide a few clear evidence that can constrain future models.

The left panels of Fig.\,\ref{Fig:sr_ba} present the variation of Ba and Sr relative to Fe as a function of metallicity in a classical manner. At first glance both [Sr/Fe] and [Ba/Fe]
present a very large scatter at any given [Fe/H]. A closer
look though reveals different behaviours, which depend on the metallicity range
and on the galaxy:  i) [Fe/H] $\simeq -2.8$ is a metallicity threshold below and above which the dispersion
in abundance ratio changes and ii) massive and fainter galaxies do not follow the same trend.

We first concentrate on our MW halo sample. In both LTE and NLTE abundance
analyses, [Sr/Fe] and [Ba/Fe] have large dispersion below [Fe/H] $\simeq -2.8$, as
also shown by \citet{2009A&A...494.1083A,2011A&A...530A.105A} and
\citet{2013A&A...551A..57H}. Above this metallicity, [Sr/Fe] becomes steadily solar.
As to [Ba/Fe], the rise to the solar value comes at slightly higher metallicity, [Fe/H] $\simeq -2.5$. Although largely diminished, the dispersion is larger than for [Sr/Fe].

\begin{figure}  
\hspace{-3mm} \resizebox{95mm}{!}{\includegraphics{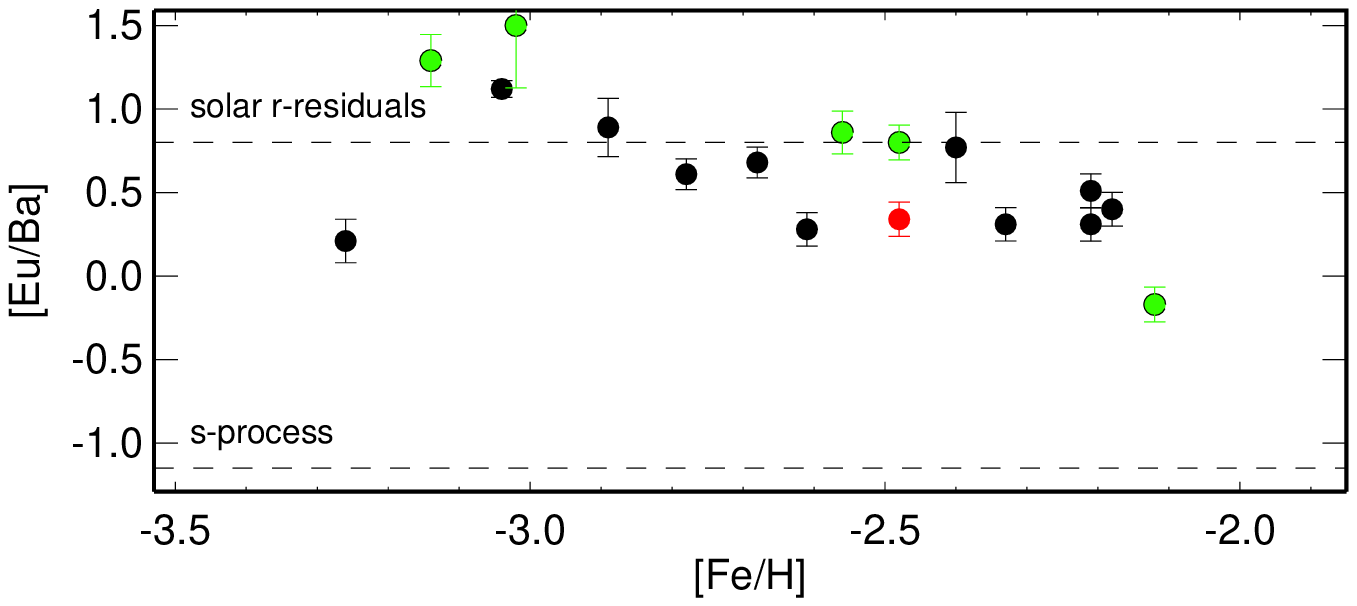}}

\vspace{-5mm}
\hspace{-3mm} \resizebox{95mm}{!}{\includegraphics{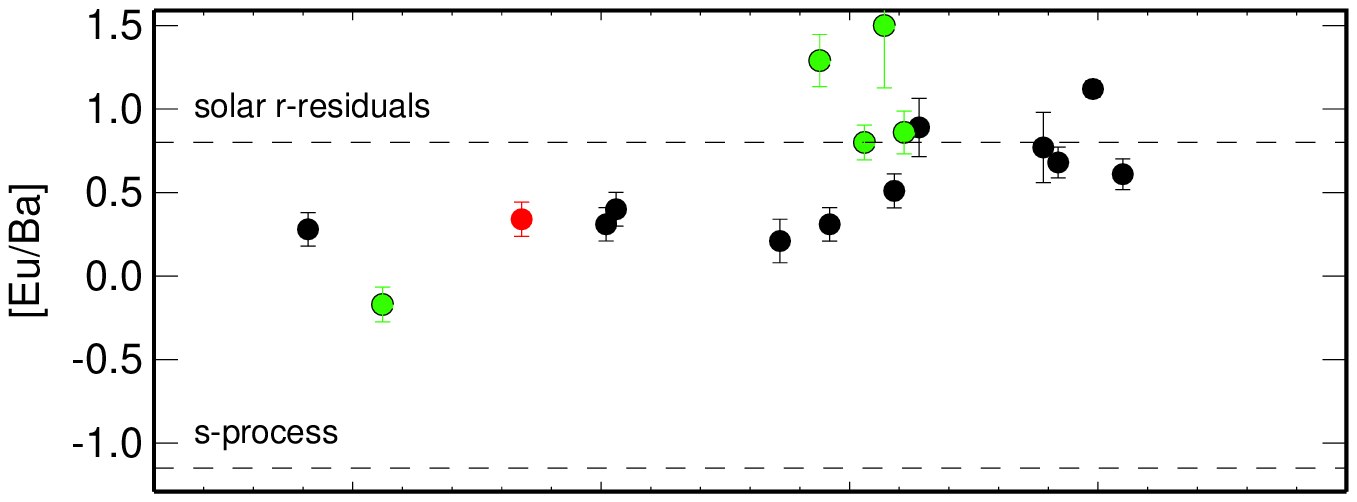}}

\vspace{-13mm}
\hspace{-3mm} \resizebox{95mm}{!}{\includegraphics{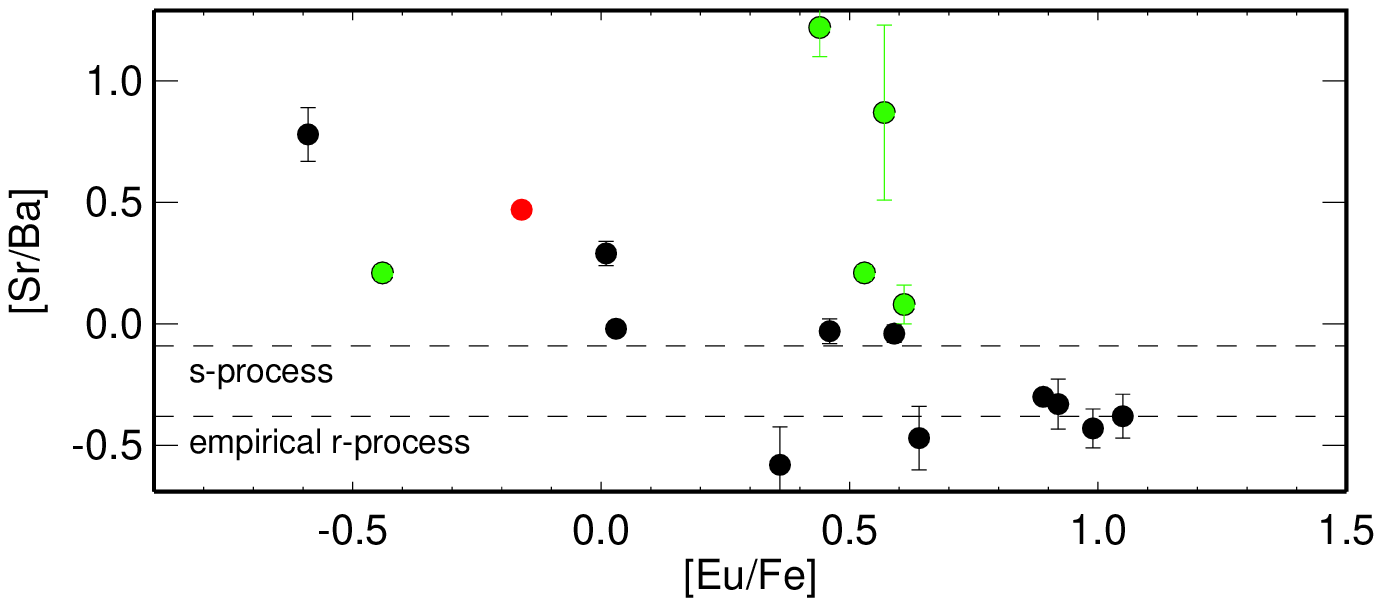}}

  \caption{\label{Fig:euba_eufe} Stellar [Eu/Ba] (top and middle panels) and [Sr/Ba] (bottom panel) NLTE abundance ratios 
in the MW halo (black circles) and the Sculptor (red circle) and Ursa Minor (green circles) dSphs. The dashed lines indicate the r- and s-process ratios [Eu/Ba]$_r = 0.80$, [Eu/Ba]$_s = -1.15$, and [Sr/Ba]$_s = -0.09$ according to \citet{2014arXiv1403.1764B} and our empirical estimate of [Sr/Ba]$_r = -0.38$.
}
\end{figure}

Europium abundances are available for 11 stars of our MW sample and, in particular, for
  all stars but HD~218857 (because its spectrum does not extend
  blue enough) at [Fe/H] $> -2.5$. Figure\,\ref{Fig:euba_eufe} shows that all
  these stars have [Eu/Ba] $\ge$ 0.28, a value much closer to the r-process
  [Eu/Ba]$_{\rm r}$ = 0.80 \citep[based on the solar
    r-residuals,][]{2014arXiv1403.1764B} than the s-process [Eu/Ba]$_{\rm s} =
  -1.15$ ratio. This reflects the fact that if there is a contribution of the
  s-process to the Ba abundances, it is only a minor one. We stress that we 
  discard the most metal-rich star of our sample, HD~8724, with [Fe/H] =
  $-1.76$ and [Eu/Ba] = 0.21, from any discussion on very metal-poor stars.

\begin{figure}  
\hspace{-3mm}   \resizebox{95mm}{!}{\includegraphics{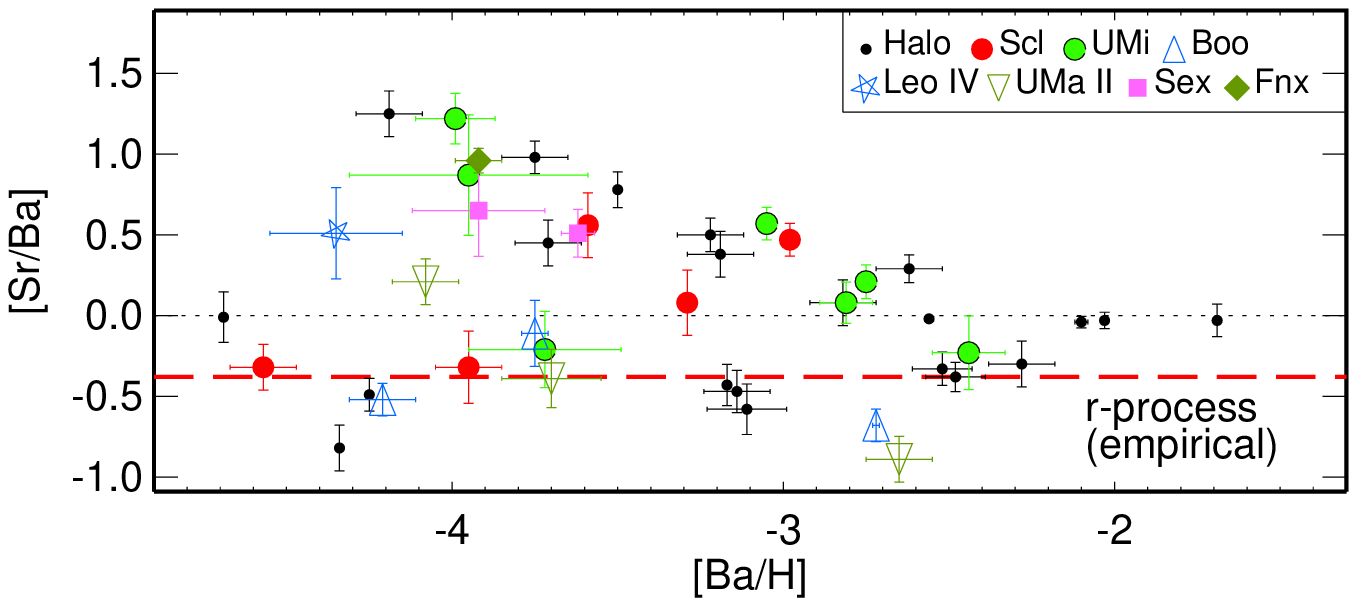}}
  \caption{\label{Fig:srba_bah} The [Sr/Ba] NLTE abundance ratios as a function of Ba abundance. The dashed line indicates the mean [Sr/Ba] ratio of the six r-II stars (see text). Symbols and colours are as in Fig.~\ref{Fig:sr_ba}. 
}
\end{figure}

Would Ba and Sr be produced by the same
nucleosynthesis source, this should result in a fairly flat (within
observational error bars) [Sr/Ba] ratio versus [Ba/H].  This is clearly not the case in Fig.\,\ref{Fig:srba_bah}.
Our MW halo sample is separated into two groups. The first one that includes 
eight of 20 stars has indeed similar [Sr/Ba] $\sim -0.5$ on the entire range of Ba abundances.
Although astrophysical site(s) of the r-process is (are) not identified yet \citep{2014ApJ...789L..39W,2017ApJ...836L..21N}, the strongly r-process enhanced ([Eu/Fe] $>$ 1, [Eu/Ba] $>$ 0) stars referred to as r-II stars \citep{HERESI} provide an observational evidence for 
 the r-process to yield a subsolar Sr/Ba ratio. We estimate the empirical r-process ratio, [Sr/Ba]$_{\rm r} = -0.38$, using the six halo r-II stars: CS~22892-052 \citep{Sneden2003}, HE1219-0312 \citep{HE1219}, SDSS~J2357-0052 \citep{Aoki2010}, HE2327-5642 \citep{HE2327}, CS~31082-001 
 \citep{2013A&A...550A.122S}, and CS~29497-004 \citep{2016arXiv160807463H}. It is worth noting that a sample of 253 metal-poor halo stars in \citet{HERESII} includes eight r-II stars and they all have the lowest and similar Sr/Ba, with the mean [Sr/Ba]$_{\rm r-II} = -0.44\pm0.08$ that is very close to our estimate of [Sr/Ba]$_{\rm r}$. The question is whether the observed subsolar Sr/Ba ratio itself can be considered as a signature of the r-process origin of Sr.

The second MW group seems to be aligned on a well-defined downward trend of [Sr/Ba] with [Ba/H].
Similar tight anti-correlation of [Sr/Ba] with [Ba/Fe] and [Ba/H] was reported by \citet{Honda2004} and \citet{Francois2007}. In line with \citet{HERESII}, we obtained that an enhancement of Sr relative to Ba correlates with the stellar Eu abundance (Fig.\,\ref{Fig:euba_eufe}): no star with supersolar Sr/Ba ratio is enhanced in Eu, while the stars with subsolar Sr/Ba ratios have [Eu/Fe] $\ge$ 0.36.

In order to explain an excess of Sr production relative to the classical r-process, various ideas and models were proposed: i) the weak s-process during the hydrostatic core He-burning phase of massive stars \citep{1991ApJ...367..228R}, ii) charged particle reactions in core-collapse supernovae \citep{1992ApJ...395..202W}, iii) nucleosynthesis from progenitor stars that lived and died prior to the formation of the first "main" r-process stars \citep{2002PASP..114.1293T}; 
iv) non-standard s-process in low metallicity massive rotating stars \citep{2008ApJ...687L..95P};
v) explosive nucleosynthesis in a high energy SN (or "hypernova") \citep{2009ApJ...692.1517I}; vi) rapid charged-particle reactions in the high-entropy winds at low entropies \citep{Farouqi2010}; 
vii) neutron star mergers \citep{2014A&A...565L...5T,2014ApJ...789L..39W}; viii) the weak r-process (referred also to as alpha-process) taking place in neutrino-driven winds \citep{2017JPhG...44e4003B}, and ix) the intermediate r-process in core-collapse supernovae driven by the magneto-rotational instability \citep{2017ApJ...836L..21N}. However, the source(s) is (are) not identified yet.

The confrontation of galaxies with very different star formation histories and
chemical evolution paths provides new abundance trends and should help to constrain
the nucleosynthesis origins of the neutron-capture elements.

We find differences in the abundance trends between the classical dSphs (Sculptor, Ursa Minor, Fornax, Sextans) and the UFDs (Bo\"otes~I, UMa~II) and, for the classical dSphs, differences between the elemental ratios involving Sr and Ba.
The main observational results can be summarised as follows.

Keeping the same distinctive two regions below and above [Fe/H] $\simeq -2.8$,
$-2.5$ for Sr and Ba, respectively, Fig.\,\ref{Fig:sr_ba} shows that the classical dSph and the MW
halo stars behave identically in the low-metallicity regions, with large
dispersions in [Sr/Fe] or [Ba/Fe] in both cases.

In the two UFDs, Bo\"otes~I and UMa~II, all the stars are 
depleted in Sr and Ba relative to Fe, with very similar ratios of [Sr/Fe] $\simeq -1.1$ and [Ba/Fe] $\simeq -0.75$ on the entire range of metallicity (Fig.\,\ref{Fig:sr_ba}).
Their Ba/Fe ratio is 
close to the Ba/Fe floor of the MW halo, while Sr/Fe is higher than that for the most Sr-poor stars in the MW, by more than 0.5~dex. The Sr/Fe ratio of the only star available in Leo~IV-S1 does not stand out of the corresponding ratios in Bo\"otes~I and UMa~II, but Ba/Fe is lower, by one order of magnitude, although remains
at the level of Ba/Fe for the MW most Ba-poor stars. Low [Ba/Fe] ratios had been reported  by \citet{2010ApJ...708..560F} and \citet{2012AJ....144..168K} based on their LTE analyses. The improvement here is that homogeneity of the analysis allows to accurately compare the levels of the stellar abundance ratios.

In order to put all galaxies on the same footing and, in particular, to remove
any potential pollution of the iron abundances by the ejecta of SNeIa that would
affect some galaxies at [Fe/H] $\ge -2.8$ but not the others, we now consider
the evolution of Ba and Sr relatively to Mg, as shown in the right panels of
Fig.\,\ref{Fig:sr_ba}. The same conclusions, as drawn below, would be reached,
should one consider any other $\alpha$-element. 

The dichotomy seen previously remains. Large scatter of data below [Mg/H] $\simeq -2.6$ in the massive galaxies confirms that their early enrichment in Ba and Sr is produced in a
variety of conditions, sufficiently rare and with varying yields for the
interstellar medium to remain inhomogeneous. A more ordered behaviour of Sr/Mg and Ba/Mg is observed above [Mg/H] $\simeq -2.6$. Strontium and barium present different features though.

For Sr/Mg, there are two clear sequences at [Mg/H] $\succsim -2.6$, one scattering
around [Mg/H] $\simeq -0.37$, which gathers the MW halo stars and the most massive
(classical) dwarfs, such as Sextans, Ursa Minor and Sculptor, while the UFDs
Leo~IV, UMa~II, and Bo\"otes~I keep remarkably consistent and low abundance
ratios, at the level of [Sr/Mg] $\simeq -1.3$. This means that while the stellar
population of massive dSphs as well as the MW halo can rise their abundance of
Sr relatively to Mg, the UFDs cannot and miss this additional Sr production channel.

As to Ba, again similar ratios, at the level of [Ba/Mg] $\sim -1$, are found in the UFDs Bo\"otes~I and UMa~II on the entire Mg abundance range. In the classical dSphs, the scatter of Ba/Mg is reduced above [Mg/H] $\simeq -2.4$, but much less than for Sr/Mg, and one does not witness any particular differential behaviour
between the classical and faint dwarfs. While the dSph stars populate [Ba/Mg]
regions where one finds the MW stars as well, it is fair to say that they do not
reach as high ratios as our MW sample stars.

Previous LTE observational investigations, comparing dwarf spheroidal galaxies with the Milky Way halo population, have shown evidence for more than one channel of production
of the neutron-capture elements lighter than Ba \citep{2010ApJ...719..931C,2010ApJ...708..560F,2010A&A...524A..58T,2012AJ....144..168K,2015A&A...583A..67J,2017ApJ...835...23R}. However, at this stage the question of the different production origins is still wide open. Our goal in the following
is to provide concrete evidence and constraints that can
be further used to improve theoretical models.

Since neither Fe nor Mg is a good tracer for the heavy or light neutron-capture
elements, we inspect the Sr/Ba ratio as a function of [Ba/H] (Fig.\,\ref{Fig:srba_bah}). 

For the UFDs, the statistics of Sr/Ba is even poorer than that for the Sr abundances. 
Five of our seven stars in the UFDs with
  such measurements have subsolar [Sr/Ba], $-0.11$ down to $-0.89$,
  suggesting a common origin of the neutron-capture
  elements in the Bo\"otes~I and UMa~II stars in the classical main r-process. The analysis of
  the r-process-rich galaxy Reticulum~II has provided evidence for a variety of evolution among
  UFDs and a reference to the r-process enhanced stars in these systems,
  which have similar [Sr/Ba] ratios of $-0.21$ to $-1.17$
  \citep{2016ApJ...830...93J,2016AJ....151...82R}. 
  
The classical dSphs follow the trends defined by the MW halo stars, either at subsolar [Sr/Ba] or along the upward [Sr/Ba] trend with decreasing [Ba/H].
In the Sculptor dSph, the upward trend extends, probably, to [Ba/H] $\sim -5$. Indeed, we have
 the two stars, Scl002\_06 and Scl074\_02, where the \ion{Ba}{ii} 4934\,\AA\ resonance line could not be measured resulting in [Sr/Ba] $> 1.1$, while their [Ba/H] $\precsim -5$. 
For our MW sample, the statistics of Sr/Ba measurements at [Ba/H] $< -4$ is poor, however, the upward [Sr/Ba] trend extending down to [Ba/H] $\simeq -5$ was reported by \citet{Honda2004}.

\begin{figure}  
\hspace{-3mm} \resizebox{95mm}{!}{\includegraphics{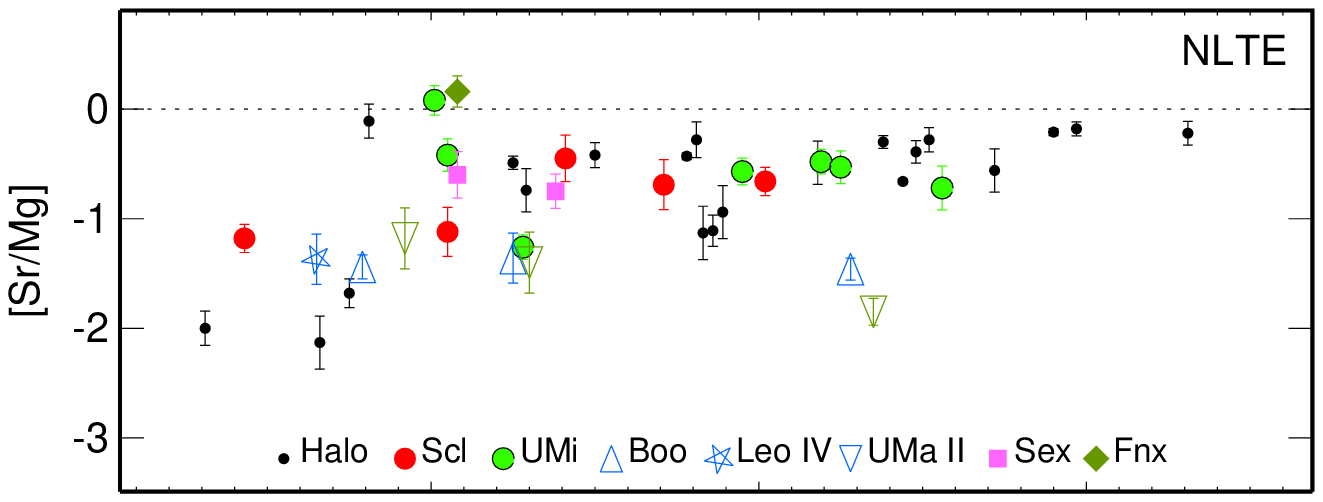}}

  \vspace{-13mm}
\hspace{-3mm} \resizebox{95mm}{!}{\includegraphics{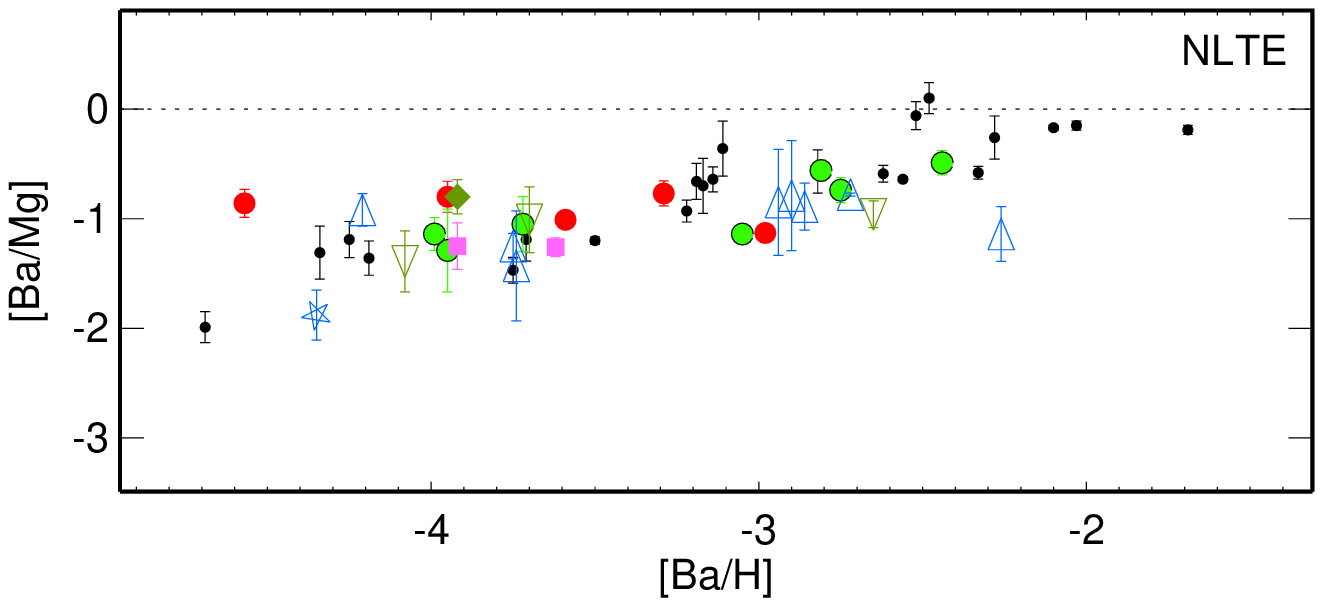}}
  \caption{\label{Fig:srbamg_bah} The [Sr/Mg] and [Ba/Mg] abundance ratios versus [Ba/H]. Symbols as in Fig.~\ref{Fig:sr_ba}.
  }
\end{figure}

The well delineated branch at [Sr/Ba] $\ge$ 0
  is built from stars of increasing Ba and stable Sr abundances. This is seen in
  Fig.\,\ref{Fig:sr_ba} (right column) and even more clearly in
Fig.\,\ref{Fig:srbamg_bah}. For these stars, Sr scales perfectly with Mg over the range
of [Ba/H] considered for the branch, implying that the second
  producer of strontium in the $-4 \precsim$ [Ba/H] $\precsim -2.5$
  regime of the MW halo and the classical dSphs is independent of the
production of barium and connected with massive stars. In the meantime, the lower panel
is very quickly independent of these.

Different mass galaxies: the MW halo, the classical dSphs in Sculptor, Sextans, Fornax, and Ursa Minor, and the Bo\"otes~I and UMa~II UFDs reveal similar [Ba/Mg] ratios in the $-4.3 \precsim$ [Ba/H] $\precsim -3.5$ range (Fig.\,\ref{Fig:srbamg_bah}), suggesting similar efficiency of Ba production via massive stars. At the higher Ba abundances, Ba/Mg grows in the massive galaxies, while remains at low level in the UFDs.

\section{Conclusions}\label{Sect:Conclusions}

The aim of this paper is to provide the community with a
  robust framework for galaxy chemical evolution studies. For that
  purpose, we have assembled a sample of galaxies with very different
  evolutionary paths, benefiting from high-resolution spectroscopy ($R
  \ge$ 25\,000) coming from a variety of sources.  In Paper~I we
  presented a homogeneous set of accurate atmospheric parameters. In
  this study, we determined the NLTE abundances of ten chemical species in the giant stars covering the $-4 <$ [Fe/H] $< -1.7$ metallicity range and
  belonging to the Milky Way halo (23 stars), as well as the Sculptor,
  Ursa Minor, Sextans, and Fornax classical dSphs and the Bo\"otes~I,
  UMa~II, and Leo~IV UFDs (36 stars). This is the first time that
  {\it all} abundances are derived under NLTE.

For each star, abundances of Na, Mg, Al, Si, Ca, Ti, Fe, Ni, Sr, and Ba were obtained, provided the corresponding lines could be measured. 
The NLTE effects on the derived abundances are found to be different in magnitude and sign, depending on the chemical species and the atmospheric parameters. 

The first major impact of a homogeneous set of
  atmospheric parameters combined with a NLTE treatment is a substantial
  reduction of the spread in abundance ratios at given metallicity,
  compared to a simple compilation of the literature data. It
  influences the abundance trends as follows.

\begin{itemize}
\item Any discrepancy in the level of the [$\alpha$/Fe]
  plateau between different $\alpha$-elements: Mg, Ca, and Ti is now
  removed. It is valid for all galaxies and particularly visible for
  the most populated ones in our sample, Sculptor, Ursa Minor, and the
  MW halo.
\item In the [$\alpha$/Fe] versus metallicity diagrams, all classical dSphs: Sculptor, Ursa Minor, Sextans, and Fornax scatter around the mean of the Milky Way halo stars, [$\alpha$/Fe] $\simeq$ 0.3, suggesting enrichment of these galaxies by the massive stars, in numbers following classical IMF. It is worth stressing that our results for the MW giants are fully consistent with the
NLTE abundances derived for the MW halo dwarfs by \citet{lick_paperII}.
\item The most dramatic effect of NLTE is found for the Na/Al abundance ratios. The LTE analysis suggests substantial overabundance of sodium relative to aluminum in all types of galaxies, at the level of [Na/Al] $\sim$ 0.8, while [Na/Al] is reduced down to $\sim$ 0.2 in NLTE.
\item For [Na/Fe], [Na/Mg], and [Al/Mg], the Milky Way and dSphs reveal indistinguishable trends with metallicity suggesting that the processes of Na and Al synthesis (carbon burning
process) are identical in all systems, independent of their mass. 
\item The relation between Ni and Fe is extremely tight, much
tighter than with any other element produced by SNeII.
\end{itemize}

The abundance trends deliver important clues on the galaxy star
formation histories. We could firmly assess the impact of SNeIa ejecta on the chemical evolution of some of the faintest dwarfs known.


\begin{itemize}
\item The Bo\"otes~I UFD reveals a decline in $\alpha$/Fe and possesses a low [$\alpha$/Fe]
      population, with consistent evidence from the
      three elements for which data are available, Mg, Ca, and Ti. This can be a signature of the SNe~Ia contribution to iron and a duration of about 1~Gyr for star formation in this galaxy. 
\item The low [$\alpha$/Fe] value of the S1 star in the Leo~IV UFD suggests that this galaxy had a long enough star formation history to be polluted by SNeIa ejecta.

\item In contrast, Ursa Major~II ($M_V = -4.2$), which
  is the faintest of our three UFDs \citep{2012AJ....144....4M} and
  has three stars that cover the $-3 <$ [Fe/H] $< -2.3$
  metallicity range, falls exactly on the [Mg/Fe] and [Ca/Fe] plateau
  formed by the MW halo stars, indicating the dominance of SNeII in
  its chemical evolution and a short formation timescale.
\end{itemize}
 
We bring in constraints on nucleosynthesis sites,
  particularly for the neutron-capture elements, and on the mixing of the
  SN ejecta in the galaxy interstellar medium and provide further
  evidence that the mass of a galaxy is an important driver of its
  chemical evolution.

 \begin{itemize}
 \item Inhomogeneous mixing 
and/or stochastic effects from small numbers of  SNe~II is robustly documented in the Sculptor dSph. One star, ET0381, is strongly deficient in all elements except
  for the Fe-group, possibly missing the ejecta of the most massive
  Type~II supernovae \citep{2015A&A...583A..67J}.
 The star 11\_1\_4296 is Mg- and Ca-poor. The two [Fe/H] $\simeq -3$ stars, 002\_06 and 074\_02, show no signature of the \ion{Ba}{ii} 4934\,\AA\ resonance line in their spectra, suggesting substantially lower Ba abundance, by more than 1~dex, compared with that of the star 03\_059 which has a similar metallicity. Neither Ba, nor Sr can be measured in the two [Fe/H] $\simeq -3.7$ stars, 11\_1\_4296 and S1020549, while the \ion{Sr}{ii} and \ion{Ba}{ii} lines are reliably detected in two other stars with almost the same metallicity and even in the EMP star 07-50.     

\item The classical dSphs behave like the MW halo with respect to [Sr/Fe]:
large dispersion below [Fe/H] $\sim -3$, but close-to-solar Sr/Fe ratio above this metallicity. Magnesium seems to be a better tracer of the Sr evolution than iron. In the investigated metallicity range, none of the classical dSphs reaches a solar Ba/Fe nor Ba/Mg ratio. 
\item Magnesium seems to be a better tracer of strontium
  than iron. Even though their [Ba/Mg] ([Ba/Fe]) increases with time
  ([Mg/H] or [Fe/H]), none of the classical dSphs reaches a solar 
  Ba/Mg (Ba/Fe) ratio.

\item The massive dSphs follow the Sr/Ba trends defined by the MW halo stars, either at subsolar value or along the declining [Sr/Ba] trend with increasing [Ba/H]. This suggests two different nucleosynthesis channels for Sr. 

\item We show evidence for that the production of Sr is independent of Ba in the $-4 \precsim$ [Ba/H] $\precsim -2.5$ regime of the massive galaxies. Nearly constant [Sr/Mg] ratios  suggest a strong link to massive stars, unlike Ba.

\item The UFDs are 
depleted in Sr and Ba relative to Fe and Mg, with very similar ratios of [Sr/Mg] $\simeq -1.3$ and [Ba/Mg] $\simeq -1$ on the entire range of metallicity (and Mg abundance) for Bo\"otes~I and UMa~II. The only star available in Leo~IV does not stand out with respect to [Sr/Mg], while its lower [Ba/Mg] ratio compared with that for the other two UFDs is probably related to the early dispersed mode of Ba production. Subsolar Sr/Ba ratios of the stars in Bo\"otes~I and UMa~II may indicate a common origin of Sr and Ba in the classical main r-process, though the statistics of Sr/Ba measurements is poor. The fact that the faint galaxies miss the second channel of Sr production could be explained by an undersampling of the IMF, possibly at its high-tail.
\end{itemize}

This study is the first step to our goal of further NLTE
  high-resolution studies of dSphs. We plan to increase the number and mass range of the galaxies explored, and to extend the observed metallicity range of their populations.
 

\begin{acknowledgements}
  
We thank Judith G. Cohen, Rana Ezzeddine, Anna Frebel, and Joshua D. Simon for providing stellar spectra, David Yong for double checking the observed equivalent widths of the \ion{Ba}{ii} lines in the Bo\"otes~I stars.
The authors are indebted to the International Space Science Institute (ISSI), Bern, Switzerland, for supporting and funding the international team "The Formation and Evolution of the Galactic Halo". 
L.M., Y.P., and T.S. acknowledge financial support from the grant on Leading Scientific Schools 9951.2016.2. 
We made use of the MARCS, SIMBAD, and VALD databases.

\end{acknowledgements}

\bibliography{mashonkina,atomic_data,nlte,scl,references,mp_stars}
\bibliographystyle{aa}

\begin{appendix}

\section{Stellar abundances}

\longtab[1]{
\begin{longtable}[]{clrrcrrcccrc}   
\caption{\label{Tab:AbundanceSummary} Summary of the LTE and NLTE abundances of the investigated stars.}  \\
\hline \noalign{\smallskip}
 Z & Atom & $N_l$ & \multicolumn{4}{c}{LTE} &  & \multicolumn{3}{c}{NLTE} & Method of \\  
\cline{4-7} \noalign{\smallskip}
\cline{9-11} \noalign{\smallskip}
   &      &               & $\eps{}$ & $\sigma_{\eps{}}$ & [X/H] & [X/Fe] & & $\eps{}$ & $\sigma_{\eps{}}$ & [X/Fe] & analysis \\
 \noalign{\smallskip}  \hline \noalign{\smallskip} 
\endfirsthead
\caption{continued.}\\
\hline \noalign{\smallskip}
 Z & Atom & $N_l$ & \multicolumn{4}{c}{LTE} &  & \multicolumn{3}{c}{NLTE} & Method of \\  
\cline{4-7} \noalign{\smallskip}
\cline{9-11} \noalign{\smallskip}
   &      &               & $\eps{}$ & $\sigma_{\eps{}}$ & [X/H] & [X/Fe] & & $\eps{}$ & $\sigma_{\eps{}}$ & [X/Fe] & analysis \\
 \noalign{\smallskip}  \hline  \noalign{\smallskip}    
\endhead
\hline
\endfoot
\hline
\endlastfoot
\multicolumn{12}{c}{Sculptor ET0381, $\Teff$ = 4570 K, log~$g$ = 1.17, [Fe/H] = $-2.19$, $\xi_t$ = 1.7\,\kms } \\  
 11 & Na I  &   2 &  3.43 & 0.11 & $-$2.90 & $-$0.71 &   &  3.13 & 0.08 & $-$1.01 & $W_{obs}$(JNM15) \\
 12 & Mg I  &   3 &  4.98 & 0.07 & $-$2.60 & $-$0.41 &   &  5.00 & 0.07 & $-$0.39 & $W_{obs}$(JNM15) \\
 13 & Al I  &   1 &  2.72 &      & $-$3.75 & $-$1.56 &   &  2.82 &      & $-$1.46 & syn \\
 14 & Si I  &   2 &  4.80 & 0.04 & $-$2.75 & $-$0.56 &   &  4.79 & 0.01 & $-$0.57 & syn \\
 20 & Ca I  &   9 &  3.73 & 0.05 & $-$2.63 & $-$0.44 &   &  3.85 & 0.05 & $-$0.32 & $W_{obs}$(JNM15) \\
 22 & Ti I  &  10 &  2.09 & 0.17 & $-$2.84 & $-$0.65 &   &  2.31 & 0.15 & $-$0.43 & idem \\
 22 & Ti II &  21 &  2.60 & 0.17 & $-$2.33 & $-$0.14 &   &  2.60 & 0.17 & $-$0.14 & idem \\
\end{longtable}
}

\end{appendix}

\end{document}